\documentclass[
  reprint,
  amsmath,amssymb,
  aps,
  prl,
  superscriptaddress,
]{revtex4-2}

\usepackage{graphicx}   
\usepackage{dcolumn}    
\usepackage{bm}         
\usepackage{amsmath}
\usepackage{tikz}
\usepackage{xcolor}
\usepackage{comment}
\usetikzlibrary{arrows.meta,patterns}
\definecolor{glassc}{RGB}{86,140,205}
\definecolor{navyedge}{RGB}{26,38,74}
\tikzset{
  glass/.style={fill=glassc,fill opacity=0.16},
  stedgeV/.style={draw=navyedge,line width=0.55pt,line join=round},
  stedgeH/.style={draw=navyedge!40,line width=0.4pt,densely dashed}}
\usetikzlibrary{calc}
\definecolor{blobblue}{RGB}{216,231,245}
\tikzset{
  blob/.style={circle, draw=black, line width=0.9pt, fill=blobblue,
               minimum size=#1, inner sep=0pt},
  leg/.style={line width=0.8pt, black!85},
  root/.style={line width=0.9pt, black},
  vtx/.style={circle, fill=black, inner sep=1.7pt},
}
\usepackage{hyperref}   
\hypersetup{colorlinks=true, allcolors=blue}

\newcommand{\pdfsecit}[1]{\pdfbookmark[1]{#1}{#1}\textit{#1}}

\DeclareMathAlphabet      {\mathit}{OT1}{cmr}{m}{it}

\begin{document}

\preprint{APS/123-QED}

\title{Surface Water Wave Scattering and the Hydrotope}

\author{Nima Arkani-Hamed}
\affiliation{School of Natural Sciences, Institute for Advanced Study, Princeton, NJ, 08540, USA}

\author{Francesco Calisto}
\affiliation{Walter Burke Institute for Theoretical Physics and Leinweber Forum for Theoretical Physics,
California Institute of Technology, Pasadena, CA 91125, USA}

\author{Nail Ussembayev}
\affiliation{School of Natural Sciences, Institute for Advanced Study, Princeton, NJ, 08540, USA}

\author{W. Wayne Zhao}
\affiliation{Department of Physics, Columbia University, New York, 10027, USA}

\author{Zihan Zhou}
\affiliation{Department of Physics, Princeton University, Princeton, NJ 08544, USA}

\date{\today}

\begin{abstract}
We study the classical tree-level scattering amplitudes of deep-water surface gravity waves using the methods of high-energy physics. For scattering in one horizontal dimension and in the two-negative-wavenumber sector we obtain a closed formula for $n$-wave scattering. Up to a kinematic prefactor, the amplitude is the volume of a classic polytope---a
box sliced by a hyperplane, which we dub the \emph{hydrotope}, whose purpose in life is simply to  organize the sign patterns of the ``chambers" characterizing all the different regions of the two-minus kinematic space. The general formula was discovered by Claude Opus 4.6 working under our guidance, beginning with our earlier discovery of a one-term expression valid in the ``simplest" kinematic chamber. Our results resolve the puzzle raised by Y.~V.~Lvov's 1997 computation of the five-wave
amplitudes, unifying and extending it to all multiplicities.
\end{abstract}

\maketitle

\pdfsecit{Introduction}---Fluid and nonlinear wave systems repeatedly reveal unexpected mathematical structure: complete
integrability and solitons~\cite{KdV}, exact turbulent cascade spectra~\cite{ZLF}, and even
topologically protected edge modes~\cite{TopoEquatorial,TongGauge}. However, deep-water surface gravity waves have largely resisted such simplification: they are nonlinear, dispersive, and spatially nonlocal. They are nonetheless of great practical consequence, governing the nonlinear energy transfer
across the ocean spectrum and the formation of rogue waves~\cite{Rogue}.
The classical framework for their nonlinear dynamics is the theory of wave
turbulence~\cite{Zakharov,ZLF,Nazarenko}, built on Zakharov's Hamiltonian formulation and
organized around interactions on the \emph{resonant manifold}, where waves obey their long-distance dispersion relations while conserving total energy and
momentum, what is referred to as on-shell kinematics in particle physics.

Perturbative tools akin to Feynman diagram techniques have long been part
of this story, including Wyld's diagrams for Navier--Stokes turbulence~\cite{Wyld} and Hasselmann's
calculation of nonlinear energy transfer across the wave spectrum~\cite{Hasselmann,hasselmann1966feynman}, but explicit interaction coefficients remain notoriously hard to come by. Nevertheless, calculations for surface waves in deep water have yielded unexpected simplicity.
In one horizontal dimension the four-wave interaction coefficients vanish identically on
the resonant manifold~\cite{DZ94,CW95}, once thought to suggest that the equations might be integrable for real kinematics~\cite{DLZ95}. Later, Lvov showed the five-wave interaction coefficients do not vanish everywhere~\cite{Lvov}. Lvov extracted the answer through a canonical transformation and summing eighty-one diagrams case by case, but the final answers proved ``remarkably simple'', a simplicity he suspected ``should have a deep physical meaning.'' What has been missing is an organizing
principle that makes such structure manifest and a formula that efficiently computes these amplitudes once and for all.

Such a principle has emerged from high-energy physics in the study of on-shell scattering amplitudes, which often prove dramatically simpler than their Feynman-diagram expansions suggest.
The Parke--Taylor formula~\cite{ParkeTaylor} collapses an avalanche of diagrams into a single
term, on-shell recursion~\cite{BCFW} builds all tree amplitudes from three-point data, and broad
classes of amplitudes are now understood as the canonical forms of \emph{positive
geometries}---the amplituhedron~\cite{Amplituhedron}, the associahedron~\cite{Associahedron},
the cosmological polytope~\cite{CosmoPolytope}---for which regions of kinematic space are carved
out by natural geometric questions, and each amplitude \emph{is} the canonical form~\cite{PosGeo}. Developed across domains as varied as gravity, gauge theory, and cosmology, these ideas invite us to ask whether similar structures lie behind non-relativistic fluid dynamics.

We find first indications pointing to a positive answer in the context of deep-water surface gravity waves. Treating each interaction as a classical scattering process we study the on-shell amplitude, a function of only external frequencies and momenta, obeying the on-shell condition. Our central result is the computation of scattering amplitudes in one spatial dimension in the leading nontrivial sector with two negative wavenumbers, an analogue of MHV amplitudes. These enjoy a closed-form formula for all multiplicities, Eq.~\eqref{eq:hydrotope}. This formula was discovered by a Claude Code AI agent (Opus 4.6) working under our guidance, beginning with our own earlier discovery of a one-term  expression Eq.~\eqref{eq:simplest}, valid in the ``simplest" kinematic chamber.

Up to a kinematic prefactor, the amplitude can be given a geometric interpretation in Eq.~\eqref{eq:geometric}, as the volume of a single polytope---a hypercube
sliced by a hyperplane---which we call the \emph{hydrotope}. This polytope has a very natural, purely kinematic origin, encapsulating all the sign-chambers of  one-dimensional kinematics. It is very striking that the volume of this kinematic object computes the scattering amplitude. 

Our formula is valid at all $n$, but up to $n=5$ points, all allowed kinematics have a single minus momentum (for which the amplitude vanishes), two minus momenta, or are related by parity to either, so our two-minus formula suffices for all five-wave amplitudes, reproducing Lvov's entire five-wave table in a single stroke. 

\vspace{6pt}

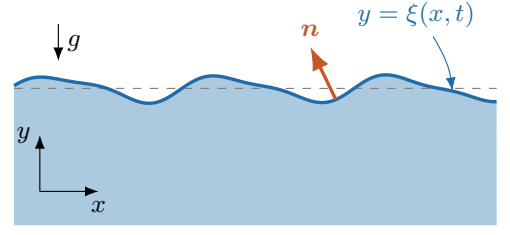
\begin{figure}[h]
\centering
\resizebox{0.75\columnwidth}{!}{%
\begin{tikzpicture}[>=Latex]
  \definecolor{waterfill}{RGB}{171,201,223}
  \definecolor{waterline}{RGB}{31,108,168}
  \definecolor{nrm}{RGB}{198,88,40}
  \def\surf{0.15*sin(\x*170)+0.045*sin(\x*330+50)}
  \fill[waterfill]
    plot[domain=0:6,samples=140,smooth] (\x,{\surf})
    -- (6,-1.7) -- (0,-1.7) -- cycle;
  \draw[gray!100,dashed] (0,0) -- (6,0);
  \draw[waterline,very thick]
    plot[domain=0:6,samples=140,smooth] (\x,{\surf});
  \draw[black,->] (0.55,0.80) -- (0.55,0.34) node[midway,right]{$g$};
  \coordinate (sp) at (4.0,{0.15*sin(4.0*170)+0.045*sin(4.0*330+50)});
  \draw[nrm,very thick,->] (sp) -- ++(-0.32,0.66) node[above]{$\bm{n}$};
  \node[waterline] (lab) at (5.0,0.92) {$y=\xi(x,t)$};
  \draw[waterline,->] (lab) to[bend left=12]
    (5.45,{0.15*sin(5.45*170)+0.045*sin(5.45*330+50)});
  \draw[black,->] (0.32,-1.28) -- (0.32,-0.58) node[left]{$y$};
  \draw[black,->] (0.32,-1.28) -- (1.05,-1.28) node[below]{$x$};
\end{tikzpicture}}
\caption{Surface gravity waves on deep water: an incompressible, irrotational, inviscid fluid (infinite depth, $h\to\infty$) with free surface $y=\xi(x,t)$, outward normal $\bm{n}$, and gravity $g$. The dashed line marks the equilibrium level $\xi=0$, and the rigid seafloor is far below in the negative $y$-direction.}
\label{fig:setup}
\end{figure}

\pdfsecit{The surface water wave problem}---
For surface water waves, start with a $d$-dimensional region occupying a
time-dependent fluid domain $\Omega(t) \subseteq \mathbb{R}^{d}_x \times \mathbb{R}^1_y$. The spatial coordinates are $(x, y)$, where $x = (x_1, \ldots, x_{d}) \in \mathbb{R}^{d}$ denotes
the horizontal directions and $y \in \mathbb{R}$ is the vertical coordinate, and gravity $g > 0$ is oriented along the negative vertical axis (see Fig.~\ref{fig:setup}).

The boundary of $\Omega(t)$ consists of two components. The bottom is described by a rigid
boundary at depth $y = -h(x)$. The upper boundary is the free surface
\begin{equation}
\Sigma(t) = \bigl\{(x, y) : y = \xi(x, t)\bigr\},
\end{equation}
where we call $\xi$ the free surface displacement, where the surface waves in question live. Deep water means we take $\min (h(x))\to\infty$, much larger than any other length scale, namely the wavelengths of the surface waves. The bulk fluid domain is then
\begin{equation}
\Omega(t) = \bigl\{(x, y) \in \mathbb{R}^{d} \times \mathbb{R} : -h < y < \xi(x, t)\bigr\}.
\end{equation}
We take the bulk fluid to be inviscid, irrotational, and incompressible, so its velocity
field derives from a scalar potential field $\phi(t,x,y)$ obeying Laplace's equation, $\nabla^2\phi+\partial_y^2\phi=0$ in the bulk.

Starting from Luke's Lagrangian formulation~\cite{Luke}, the action is
\begin{equation}\label{eq:bulkaction}
\begin{split}
S[\phi,\xi]=-\rho\!\int\! dt\,d^d x\!\int_{-h(x)}^{\xi(x,t)}\!\!\! dy\,&\bigl[\,
\partial_t\phi+\tfrac12(\nabla\phi)^2\\[-2pt]
& +\tfrac12(\partial_y\phi)^2+gy\,\bigr];
\end{split}
\end{equation}
which recovers Laplace's equation under variation of $\phi$, which is exponentially suppressed at the bottom boundary $y=-h\to -\infty$ in deep water. All the nontrivial part of this action lives at the upper boundary, the dynamical free surface $\xi(x,t)$. Varying the fields at $y=\xi(x,t)$ gives the two equations of motion at the upper boundary:
\begin{align}
    & \partial_t\xi+\nabla\xi\cdot(\nabla\phi)_\xi-(\partial_y\phi)_\xi=0 \label{eq:skc} \\
    & (\partial_t\phi)_\xi+\tfrac12(\nabla\phi)_\xi^2+\tfrac12(\partial_y\phi)_\xi^2+g\xi=0 \ ,
\end{align}
with $\phi$ and its derivatives evaluated at $y=\xi(x,t)$ in both equations, e.g. $(\phi)_\xi = \phi(t,x,y)_{y=\xi}$, $(\nabla\phi)_\xi = (\nabla_y\phi(t,x,y))_{y=\xi}$. The first is a \textit{kinematic} condition encoding that the surface is comoving with the bulk fluid, and the second is a \textit{dynamic} condition, Bernoulli's equation for unsteady potential flow evaluated on the surface.  The symmetries and conservation laws of this problem have been studied in~\cite{BenjaminOlver}.

For surface gravity waves, we would like to find a surface action described by purely surface variables. To this end, we can integrate out the bulk leaving
\begin{equation}\label{eq:bulkactionintegrated}
S[\phi,\xi]=\frac{\rho}{2}\!\int\! dt\,d^d x\left[(\phi)_{\xi} \partial_t\xi - g\,\xi^2\right]
\end{equation}
The hydrostatic term integrates over the water column to the gravitational potential energy $\tfrac12 g\,\xi^2$, giving the restoring potential. For concreteness, we have chosen to work in a regime where we neglect surface tension in this work, but there is no obstruction to including surface tension in this formalism.

Now, since we are literally scattering gravity waves of surface displacement $\xi$, we would like to express the entire action in terms of $\xi$ alone. We do this by inverting \eqref{eq:skc}, trading Dirichlet data $(\phi)_\xi$ (also called $\psi$ in the literature) for Neumann data $\nabla \xi$. This inversion is solved with a perturbative expansion in a small dimensionless parameter called surface steepness $\epsilon \sim\xi/\lambda\sim k\xi$, the aspect ratio of the vertical height of a wave to its horizontal wavelength $\lambda$. Density $\rho$ is a constant that completely factors out and we can work in units where gravitational acceleration $g=1$, so $\epsilon$ is the remaining small ($\lesssim 0.4$) parameter in the problem.
We end up with an effective field theory Lagrangian completely specified by the surface displacement $\xi$, in spatial momentum space~\footnote{Here, expressions such as $(\partial_t \xi)^2\xi$ are a shorthand for the momentum space integrals with the accompanying momentum conserving delta functions:
\begin{equation*}
\int \tfrac{d^dk_1}{(2\pi)^d} \tfrac{d^dk_2}{(2\pi)^d} \tfrac{d^dk_3}{(2\pi)^d} \partial_t \xi (k_1)\partial_t \xi (k_2)\xi(k_3) (2\pi)^{d} \delta^{d}(k_1+k_2+k_3)
\end{equation*}}:
\begin{equation}\label{eq:surfL}
\mathcal L =  \frac{1}{2|k|}(\partial_t \xi)^2- \tfrac12 g\,\xi^2+ V_3 (\partial_t \xi)^2\xi + V_4 (\partial_t \xi)^2\xi^2+\cdots
\end{equation}
where the unsymmetrized contact terms can be found in~\cite{Hydrotopecode}.
Fourier transforming time as well to get angular frequencies, we have up to cubic order,
\begin{align}\label{eq:surfLcubic}
\mathcal L = &\int d \omega\, d^d k\frac{1}{2}\left(\frac{\omega^2}{|k|}-g\right)\xi(-\omega,-k)\,\xi(\omega,k)\\
+&\int \left(\prod_{i=1}^3d\omega_i\, d^d k_i\right) (2\pi)^{d+1} \delta\left(\sum_i \omega_i\right)\delta^{d}\left(\sum_ik_i\right)\nonumber\\
&\times \frac{\omega_1\omega_2}{2}\left(\frac{k_1\cdot k_2}{|k_1||k_2|}+1\right)
\xi(\omega_1,k_1)\xi(\omega_2,k_2)\xi(\omega_3,k_3) \nonumber
\end{align}
We can read off the deep-water dispersion relation and the retarded propagator from the quadratic pieces:
\begin{equation}\label{eq:prop}
\omega^2=g\,|k|,\qquad
D(\omega,k)=\frac{|k|}{(\omega+i0^+)^2-g|k|},
\end{equation}

For the unsymmetrized three-point contact term, we generically have no poles, simply $\omega_1\omega_2 (1+\hat k_1 \cdot \hat k_2)$. This is special to three-points; at higher points generically there are poles in the contact terms, which can be accessed in the Supplemental Material and GitHub repository~\cite{Hydrotopecode}.

Working with Lagrangian contact terms, we can compute the classical field theoretic scattering amplitude by computing tree-level Feynman diagrams. As long as we are careful about the complex analytic continuation of the $\omega$ and $k$ subject to the on-shell condition $\omega^4=k^2$, we can immediately read off meromorphic Feynman rules for each vertex for positive and negative frequencies, without need to keep track of the flow of time as in the Hamiltonian formulation of \cite{Zakharov,CraigSulem,Ussembayev}; we get this for free by analytical continuation. A practical consequence is illustrated by calculation of the five-point amplitude, for which the Hamiltonian calculation requires upwards of 81 terms, whereas the Lagrangian Feynman diagram calculation only has 26. This is further improved with Berends-Giele recursion made possible by the Lagrangian vertices, discussed below.

We leave the extensive treatment of the water wave problem in general dimensions and of the symmetry breaking pattern and its consequences to upcoming work~\cite{LongPaper}, and focus on one spatial dimension in what follows.

\vspace{6pt}

\pdfsecit{One-dimensional kinematics}---
We now specialize to one horizontal dimension, $d=1$, and set $g=1$.
One-dimensional momenta are signed scalars, and because every propagator and vertex depends on momenta through absolute values of sums $\left|\sum_i k_i\right|$, the amplitude is
analytic in the external momenta only inside regions where the sign of each relevant subset sum $\sum_i k_i$ is fixed. 
For real kinematics, a sign flip occurs exactly where the absolute value kinks, leading to non-analytic amplitudes.
These regions---the \emph{kinematic chambers}---are finite and combinatorial, and the analytic structure becomes sharp. 
We need only one dynamical input: the behavior of the amplitude when a single leg goes soft, $\omega_i^2=|k_i|\to 0$.

We label one-dimensional kinematics as follows: external legs carry frequencies $\omega_1,\dots,\omega_n$ and momenta
$k_i=\sigma_i\,\omega_i^2$ with $\sigma_i=\pm1$ encoding the direction of the wavenumber, using the all-incoming convention uniformly for all legs. Their conservation equations are 
\begin{equation}
\sum_{i=1}^n\omega_i=0,\qquad
\sum_{i=1}^n k_i = \sum_{i=1}^n\sigma_i\,\omega_i^2=0.
\label{eq:conv}
\end{equation}
It must be stressed that the physical process is always $(n-1)$ linear waves in and one wave out, regardless of the signs of the $\omega_i$, $k_i$, or the incoming sign conventions.

We organize the computation by the sign pattern of the one dimensional momenta $\{\sigma_i\}$. Due to momentum conservation, all-plus and all-minus are not kinematically supported so the amplitude is zero. Moving on to the one-minus and mostly plus sector, e.g. $\sigma_1=-1$ and all other $\sigma_{i\ge2}=+1$, the amplitude vanishes for all
$n$~\cite{Ussembayev}, and the same for one-plus and mostly minus. The first nontrivial case, and the main subject of this Letter, is the \emph{two-minus} sector,
\begin{equation}
\sigma_1=\sigma_2=-1,\qquad \sigma_i=+1\ \ (i\geq3),
\end{equation} The two-plus sector is related by parity invariance of the Lagrangian. 
Detailed analysis of the three-minus sector only starts to matter at six points and higher and will be investigated in future work~\cite{LongPaper}.

\vspace{6pt}

\begin{figure}[h]
    \centering
\includegraphics[width=1.0\columnwidth]{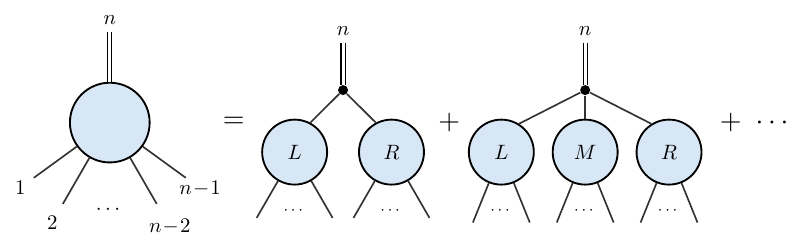}
    \caption{Berends--Giele recursion for the current $J_n$ with $(n-1)$ on-shell legs and one off-shell leg: the off-shell Lagrangian vertices at cubic and higher order partition it into sub-currents, \emph{summed over all partitions} $\{L,R\}$, $\{L,M,R\},\dots$ of the on-shell legs.}
    \label{fig:bg}
\end{figure}

\pdfsecit{Classical solutions and the single-minus sector}---
Let us spell out the physical meaning of
the classical scattering amplitude $A_n$ and of the one-leg off-shell current $J_n$ in this context. In general, the solution to the nonlinear water wave equation can be constructed by preparing $n-1$ incoming on-shell ``linear'' waves that are free and reading off the out-going ``nonlinear'' wave.
The evolution is fully causal, hence the use of the retarded propagator in Eq.~\eqref{eq:prop}. 
We note however that this prescription does not impact tree level scattering amplitudes, and it would enter only in reconstructing the causal time-domain profile.
The product of the wave heights of the incoming waves times the scattering amplitude essentially gives us the wave height of the out-going wave. 
The one-leg off-shell current $J_n$ that enters the Berends-Giele recursion pipeline \cite{BerendsGiele1988} does not have that leg amputated, and it physically represents the full off-shell ``bound'' field response  driven by $n-1$ on-shell incoming waves (see Fig.~\ref{fig:bg}).  

The amplitude is recovered by amputating the off-shell leg in the Berends-Giele current, and then taking the on-shell limit of that leg. That means explicitly $A_n = \lim_{\omega^4\to k^2}(\omega^2/|k|-1) J_n$.

The distinction between the two can be sharply illustrated in the single-minus sector. With a single minus leg and the rest positive, momentum
conservation gives every subset sum a definite sign, so there is a single chamber and the
amplitude is one analytic function.
From explicit computation, we find the amplitude $A_n(-+\cdots+)$ vanishes at all points, which is a consequence of the recursion relations studied in~\cite{Ussembayev}.

The one-leg off-shell Berends-Giele current with the other $|S|$ legs on-shell is
\begin{equation}\label{eq:Jcurrent}
J(S)= (k_S)^{\,|S|-1}\,,\qquad k_S=\sum_{j\in S}k_j\,,
\end{equation}
with $S\cup\{-k_S\}$ a set of mostly positive momenta with exactly one minus. That is, either the off-shell leg $-k_S$ is the single minus or one of the $k_{i\in S}$ is the single minus.
Since this is a solution to all multiplicities, the classical solution sourced by the one-minus currents can be explicitly found. As will be discussed in more detail in \cite{LongPaper}, a special case of this solution recovers the classical solution in \cite{ussembayev2019exactsolutionprogressivegravity}. Because $J(S)$ in these sign configurations has no poles, we can readily take the on-shell limit of the unamputated current to recover the vanishing one-minus amplitudes.

\pdfsecit{The hydrotope and the two-minus sector closed formula}---Take the two negative momenta $k_1=-\omega_1^2$ and $k_2=-\omega_2^2$ in two-minus $n$-point scattering and let $\beta^2\equiv\min(\omega_1^2,\omega_2^2)$. 
Cutting an internal line of any tree diagram separates leg-$1$ together with a subset $S$ of the plus legs from the rest, and the line carries
\begin{equation}
q_S=k_1+\sum_{i\in S}k_i=-\omega_1^2+k_S.
\label{eq:qS}
\end{equation}
The same line, cut from the leg-$2$ side, carries $k_2+k_{\bar S}=-q_S$ by momentum conservation, so the signs of the $q_S$ already fix the sign of every internal momentum: tracking $\mathrm{sgn}(q_S)$ alone labels the
``chambers'' of the kinematic space. Since $\omega^2=|k|$ is non-analytic wherever an internal momentum changes sign,
the amplitude is piecewise polynomial, with one chamber per sign pattern.

Interestingly, all of these chambers are succinctly captured by a very simple polytope geometry. Attach a coordinate $t_i\in[0,\omega_i^2]$ to each
plus leg and form the scaled hypercube $\prod_{i=3}^{n}[0,\omega_i^2]$. Its $2^{\,n-2}$ vertices are
labelled by the subset $S$ of plus legs grouped with leg $1$ (those with $t_i=\omega_i^2$)
and sit at height $\sum_i t_i=k_S$, the plus-leg subset sum. By Eq.~\eqref{eq:qS} a vertex
lies above or below the plane $\sum_i t_i=\beta^2$ according to $\mathrm{sgn}(q_S)$; knowing
which vertices fall below this single lower plane fixes the chamber, since the planes
$\sum_i t_i=\omega_1^2$ and $\sum_i t_i=\omega_2^2$ carry the same information by the
complementarity above. The scaled hypercube sliced by this plane is the \emph{hydrotope}:
\begin{equation}
\mathcal{W}_n=\Bigl\{(t_3,\dots,t_n):\,0\leq t_i\leq\omega_i^2,\ \sum_{i=3}^{n}t_i=\beta^2\Bigr\} .
\label{eq:polytope}
\end{equation}
Our central result is that
the two-minus amplitude is, up to a kinematic prefactor, the volume of the hydrotope,
\begin{equation}
A_n^{(--+\cdots+)}=\omega_1\omega_2\,2^{\,n-1}(n-3)!\,\mathrm{Vol}(\mathcal{W}_n),
\end{equation}
where we defined
\begin{equation}
\mathrm{Vol}(\mathcal{W}_n)=\!\int_0^{\omega_3^2}\!\!\!\!\cdots\!\int_0^{\omega_n^2}\!
\Big(\prod_i dt_i\Big)\,\delta\Big(\sum_i t_i-\beta^2\Big),
\label{eq:geometric}
\end{equation}
and equivalently the slice is $\partial_{\beta^2}$ of the solid region $\sum_i t_i\le\beta^2$.
Evaluating it gives the closed form at once. Fourier-representing the hyperplane constraint decouples the $t_i$ into independent edge
integrals $\int_0^{\omega_i^2}\!dt_i\,e^{i\alpha t_i}=(1-e^{i\alpha\omega_i^2})/(-i\alpha)$,
whose product expands over the $n-2$ legs, $\prod_i(1-e^{i\alpha\omega_i^2})=\sum_S(-1)^{|S|}e^{i\alpha k_S}$. 
The remaining frequency
integral at each term is a pure truncated power,
$(n-3)!\!\int\!\frac{d\alpha}{2\pi}(-i\alpha)^{-(n-2)}e^{-i\alpha P}=[P]_+^{\,n-3}$, so that
\begin{equation}
A_n^{(--+\cdots+)}=\omega_1\omega_2\,2^{\,n-1}\!\!\sum_{S\subseteq\{3,\dots,n\}}\!\!
(-1)^{|S|}\Bigl[\beta^2-\sum_{i\in S}\omega_i^2\Bigr]_+^{\,n-3},
\label{eq:hydrotope}
\end{equation}
where $\left[x\right]_+=x \Theta(x) =\max(0,x)$. This is an ordinary inclusion--exclusion formula: from the full simplex
$\{t_i\ge0,\ \sum_i t_i=\beta^2\}$ subtract each over-filled edge $t_i>\omega_i^2$, add back
the doubly-counted overlaps, and so on---one alternating term per $(n{-}2)$-cube vertex $S$,
contributing only when $k_S<\beta^2$.

There is an especially simple limit of this expression, when one of the minus legs is softest, i.e. where one of $|\omega_{1,2}|$ is the smallest of all the $|\omega|$. In this case the cut plane is very close to the corner at the origin, and the cut geometry is just a simplex. In this special limit, we have 
\begin{equation}
\left.A_n^{-- + \cdots +}\right|_{|\omega_1| \, {\rm is\,  smallest}} = 2^{n-1} \omega_1^{2n - 5}\omega_2
\label{eq:simplest}
\end{equation}
\begin{figure*}[t]
\centering
\includegraphics[width=0.84\textwidth]{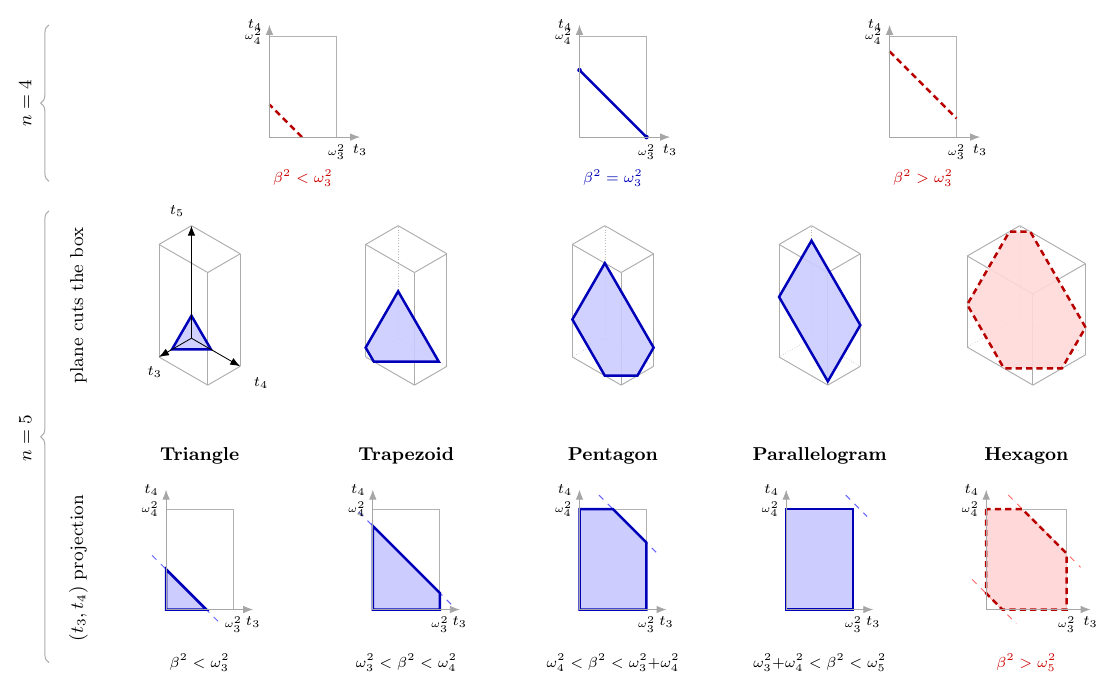}
\caption{The hydrotope $\mathcal{W}_n$ as a hyperplane section of the scaled $(n{-}2)$-cube, and manifests the kinks in subset sums of momenta appearing in the amplitude. \emph{Top
($\mathit{n=4}$):} the rectangle $[0,\omega_3^2]\times[0,\omega_4^2]$ ($\omega_3^2 \leq \omega_4^2$) cut by $t_3+t_4=\beta^2$ gives $\mathcal{W}_4$, a segment whose length is the amplitude. Energy and momentum conservation further constrains $\mathcal W_4$ to $\beta^2=\omega_3^2$. \emph{Middle and bottom
($\mathit{n=5}$):} the rectangular prism $[0,\omega_3^2]\times[0,\omega_4^2]\times[0,\omega_5^2]$ ($\omega_3^2 \leq \omega_4^2 \leq \omega_5^2$) cut by
$t_3+t_4+t_5=\beta^2$ gives $\mathcal W_5$ (middle), and the same slice in the $(t_3,t_4)$ projection (bottom). As
$\beta^2$ grows the section sweeps through the shapes shown; red marks the kinematically
forbidden regions.}
\label{fig:sgallery}
\end{figure*}
The general expression Eq.~\eqref{eq:geometric} makes central properties of the amplitude obvious. It is of course manifestly permutation symmetric in the two minus legs, and all the plus legs. It also trivially manifests the soft limit properties. As a single minus leg frequency $\omega_s$ goes to zero, the ``cutting plane" gets close to the origin and the amplitude vanishes as $(\omega_s^2)^{(n-3)}$ after factoring out $2^{n-1}\omega_1\omega_2$, compatible with the fact that the single-minus amplitudes vanish. Instead, as a plus leg goes soft, the integral manifestly vanishes linearly in $\omega_s^2$, and the coefficient of the linear term is precisely the one lower dimensional volume.\\

To reiterate, it is important to emphasize that the volume expression, up to overall normalization, is in fact entirely determined by the obvious properties expected of the amplitude together with the soft behavior. Given that the single-minus amplitudes vanish, we know that the two-minus amplitudes can not have poles. We also know that they can not be analytic functions of the momenta everywhere, for the obvious reason that {\it magnitudes} of subset sums of momenta, $|\sum k|$, occur everywhere in the computations.
And we expect the amplitudes to vanish as $\omega_s^2$ in the soft limit. As we have said these are all properties of the volume formula, and indeed if we reverse the logic, we can uniquely determine the volume function in this way. We are looking for expressions that are piecewise polynomial, with jumps precisely where $|\sum k|$ changes sign, which must vanish at least as fast as $\omega_s^2$ as any leg becomes soft. Standard results from ``spline theory" going back to the 1980's~\cite{deBoorHollig82,BoxSplines} then prove that the ``cut box volume" formula is the unique expression (up to overall numerical normalization) satisfying these properties.  

The volume formula also extends the soft behavior to an exact statement away from the soft limits, giving us that the derivative with respect to any momentum of the $n$-point volume is determined by the $(n-1)$-point volume as: 
\begin{equation}
\partial_{\omega_n^2} {\rm Vol}({\cal W}_n)={\rm Vol}({\cal W}_{n-1})|_{\omega_{1,2}^2 \to \omega_{1,2}^2 - \omega_n^2}
\end{equation}
Of course this relation, together with vanishing in the soft limit, can be inductively integrated back to the full integral form of the volume. 
  
\vspace{6pt}

\pdfsecit{Examples of the geometry}---%
As anticipated, the hydrotope has a transparent geometric meaning, which we now make explicit at four, five and six points.
The simplest case is $n=4$: given the box $[0,\omega_3^2]\times[0,\omega_4^2]$,
$\mathcal{W}_4$ is the one-dimensional segment cut by $t_3+t_4=\beta^2$ (Fig.~\ref{fig:sgallery},
top row). Here energy and momentum conservation pin the on-shell kinematics to the single wall
$\beta^2=\omega_3^2$, the degenerate base case where the four-wave amplitude vanishes~\cite{DZ94,CW95}.
The genuine multi-chamber structure begins at $n=5$. Expanding Eq.~\eqref{eq:hydrotope} gives
\begin{align}
A_5={}&16\,\omega_1\omega_2\Bigl[\beta^4
-[\beta^2-k_3]_+^2-[\beta^2-k_4]_+^2-[\beta^2-k_5]_+^2\nonumber\\
&+[\beta^2-k_3-k_4]_+^2+[\beta^2-k_3-k_5]_+^2\nonumber\\
&+[\beta^2-k_4-k_5]_+^2-[\beta^2-k_3-k_4-k_5]_+^2\Bigr],
\label{eq:n5}
\end{align}
with $k_i=\omega_i^2$. On shell $k_3+k_4+k_5=\omega_1^2+\omega_2^2\ge2\beta^2$, so the last term
vanishes and $A_5=16\,\omega_1\omega_2\,\mathcal{P}$ with $\mathcal{P}$ the square bracket above. As
$\beta^2$ grows, $\mathcal{W}_5$ runs through the cross-sections of a rectangular prism, see Fig.~\ref{fig:sgallery} with ordering $\omega_3^2\le\omega_4^2\le\omega_5^2$: a triangle
cutting off one corner, a trapezoid, a pentagon, and---when
$\omega_5^2>\omega_3^2+\omega_4^2$---a parallelogram. The one section that never occurs
is the hexagon ($\beta^2>\omega_5^2$): momentum conservation caps
$\beta^2=\min(\omega_1^2,\omega_2^2)\le\tfrac12\sum_{i=1}^2\omega_i^2=\tfrac12\sum_{i=3}^5\omega_i^2$ and energy conservation
sharpens this to $\beta^2<\omega_5^2$.

Lvov~\cite{Lvov} computed the same physics using the Hamiltonian framework: his \emph{effective five-wave Hamiltonian} matrix elements $T^{k_1k_2k_3}_{pq}$ on the resonant manifold are the physical five-wave interactions. Fixed instead by Lagrangian vertices, our on-shell amplitude reproduces his case-by-case analysis exactly. They are related by normalization,
\begin{equation}
A_5^{\,\mathrm{BG}}=\frac{16\pi^{3/2}}{\sqrt{\omega_{k_1}\omega_{k_2}\omega_{k_3}\omega_p\omega_q}}\;
T^{k_1k_2k_3}_{pq},
\end{equation}
but we bypass the construction entirely: the amplitude is a single object computed directly from on-shell data, as encoded by the closed $n$-point hydrotope formula.

For $n=6$ the section is a three-dimensional polytope---a slice of
the scaled four-cube (Fig.~\ref{fig:s6}). Sweeping $\beta^2$ and the orderings of the
$\omega_i^2$ now produces 14 distinct sectors, realizing 12
combinatorially distinct polyhedra: from the tetrahedron,
through a triangular prism, a parallelepiped, and a truncated tetrahedron, up to
twelve-vertex forms.

\vspace{6pt}

\pdfsecit{Conclusion and outlook}---
The hydrotope joins a growing web of connections between geometry and scattering amplitudes: in every
example the geometry is carved out directly by some natural question in kinematic space, yet its volume or canonical form computes a dynamical observable. The hydrotope's primary purpose is purely kinematic---it is nothing more than the simplest object that organizes all the chambers of one-dimensional kinematics at once---so that its volume should compute the scattering amplitude is, to us, the most surprising part of the story.

It is tempting to ask whether this analytic structure has any bearing on the physics of extreme surface events. Rogue waves---the localized, anomalously large events whose generation mechanism remains
debated~\cite{Rogue,Dysthe}---are only weakly nonlinear, with carrier steepness of order $\epsilon\sim\xi/\lambda\sim 0.1$~\cite{OnoratoExp} (or locally near the breaking limit, $\lesssim 0.4$~\cite{Slunyaev}), so a perturbative description based on scattering amplitudes is not obviously precluded. Our amplitudes may be used to build nonlinear wave solutions, constructing water wave profiles starting from given boundary conditions. The poles of the off-shell current seed localized extremes, and poles of the on-shell amplitude yield steeper nonlinear waves. These two phenomena, that occur at sufficiently high multiplicity, are calculable and may contribute appreciably to nonlinear rogue-wave formation. Concretely connecting the singularities of tree amplitudes to the formation of coherent extreme events is an important question we leave to upcoming work.

More broadly, the goal is to understand all amplitudes, in every sign sector and in higher dimensions, and to ask whether the structures that organize relativistic theories have water wave counterparts: are there simple analogues of the MHV expansion or BCFW-type recursion? And, ultimately, is there a unifying geometrical object for water waves? We intend to follow up this story including connecting the symmetries of the surface theory and position space wave solutions to the amplitudes in all sign sectors and higher dimensions~\cite{LongPaper}.

\vspace{6pt}

\pdfsecit{On the use of AI}---
Given the great current interest in the capabilities and use of AI in theoretical research, it is worth giving a clear picture of the nature of our (very successful) interaction with AI here.

To set the scientific context, this project is an offshoot of an effort we have been undertaking for some time, to understand the physics of water wave scattering amplitudes 
from the modern perspective of high-energy physics, especially using the ``on-shell" point of view. Most of our focus has been on the physics for general $d$-dimensional surfaces, with $d=2$ of direct relevance for real ocean waves. But obviously great simplifications always occur in $d=1$ dimensional scattering in any system, given  
extremely restricted kinematics. There were also already a number of indications for simplicity in the specific case of water waves, beginning with the observation of the vanishing of single-minus amplitudes and the relatively simple results of Lvov at five points. We therefore decided to restrict our formulae to one dimension for the two-minus configurations. The results reported in this letter were all derived in the span of less than two days. 

We first quickly found the one-term formula Eq.~\eqref{eq:simplest} valid in the ``simplest" chamber, and initially thought it might be correct globally, before finding other chambers. At this point we decided this was a perfect test case for using AI to ``discover a simple formula", since we suspected that one existed and also had a first guess for it. This was the information that was given to Claude Code, together with the Lagrangian and the BG recursion relations for computing the amplitude. At first, the agent was also not able to see past the simplest chamber, but after some prompting from us on the importance of understanding other chambers, it found the inclusion-exclusion formula of Eq.~\eqref{eq:hydrotope}.

In the first instance, then, our experience can be counted as another in a list of  examples of the striking ability of AI to ``find a simple formula", given the ability to check it against the ground truth 
(in our case BG recursion for the amplitude). Our example bears some similarity to the ChatGPT discovery of the ``single-minus" gluon amplitude 
in one-dimensional collinear kinematics~\cite{Guevara:2026qzd}. Indeed the formulae themselves are very similar, both involving inclusion/exclusion sums---apparently AI models are especially good at discovering formulae for amplitudes in one dimension! In our case, the final expression 
is so simple that ordinary human pattern recognition methods would 
almost certainly have eventually found the result---much more intricate ``simple formulae" have been discovered in standard ways. But of course it is not obvious ahead of time whether simplifications are possible, and so the ability of AI to quickly ``find a simple formula"---when one exists---is incredibly useful in the process of discovery.

Happily in our case, ``the simple formula" was so ubiquitous in the mathematical literature on splines, that the AI could also immediately tell us that the
inclusion/exclusion formula could be interpreted as a geometric volume of a box cut by a plane. We could then easily recognize the physics-reason for the existence of the geometry as described above---a polytope that captures all the possible kinematic chambers. This then serves as a springboard for looking for more general, ``kinematic geometries" that may be relevant for all the amplitudes in this theory, even if they do not all admit ``simple formulae".  Thus our experience also shows a  second use of AI---plausibly connected to its ability to discover the simple formula to begin with---in more powerfully exposing the unity of knowledge and similarities between equations across different fields that rarely intersect each other, here connecting the physics of water wave scattering amplitudes to the mathematics of splines. 

\vspace{4pt}

\pdfsecit{Acknowledgements}--- 
We thank Miguel Onorato for discussions on rogue waves and Christian Jepsen for useful comments on a draft of this work.  Z.Z. especially thanks Matias Zaldarriaga for organizing
the AI term at the IAS and for his support in facilitating the use of Claude Code, which made this discovery possible.
W.Z. is supported by the Columbia A\&S Faculty Research Allocation Program.
N.A.H. is supported by the DOE (Grant No. DE-SC0009988), the Simons Collaboration on Celestial Holography, the ERC UNIVERSE+ synergy grant, and the Carl B. Feinberg cross-disciplinary program in innovation at the IAS.
F.C. is supported by the Department of Energy (Grant No. DE-SC0011632), the Walter Burke Institute for Theoretical Physics, and the Leinweber Forum for Theoretical Physics. 

\bibliographystyle{apsrev4-2}
\bibliography{references}
\onecolumngrid
\clearpage
\begin{center}
\textbf{\large Supplemental Material}
\end{center}

\setcounter{equation}{0}\renewcommand{\theequation}{S\arabic{equation}}
\setcounter{table}{0}\renewcommand{\thetable}{S\arabic{table}}
\setcounter{figure}{0}\renewcommand{\thefigure}{S\arabic{figure}}
\renewcommand{\theHequation}{S\arabic{equation}}
\renewcommand{\theHtable}{S\arabic{table}}
\renewcommand{\theHfigure}{S\arabic{figure}}

\section{Lagrangian kernels}

Integrating out the bulk leaves a Lagrangian for the surface elevation $\xi$ alone.
With $u\equiv\dot\xi$ and the Dirichlet--Neumann operator $G(\xi)$, defined by
$u=G(\xi)\,\psi$, it reads
\begin{equation}
\mathcal{L}=\tfrac12\,u\,G(\xi)^{-1}u-\tfrac12\,g\,\xi^2 .
\label{eq:Slag}
\end{equation}
All of the nonlinearity sits in the inverse operator $G(\xi)^{-1}$: expanding it in
powers of the surface steepness defines the \emph{Lagrangian kernels} $F_n$ as the
coefficient of the interaction with two time-derivative legs (momenta $k_1,k_2$) and
$n-2$ undifferentiated legs,
\begin{equation}
\tfrac12\,u\,G(\xi)^{-1}u
=\frac{1}{2|k|} \dot \xi_{k_1}\dot \xi_{k_2}-\frac12\sum_{n\geq 3}F_n(k_1,\dots,k_n)\,\dot\xi_{k_1}\dot\xi_{k_2}\,\xi_{k_3}\!\cdots\xi_{k_n} ~.
\label{eq:Skernels}
\end{equation}
The $F_n$ are the only dynamical input; the rest of this section computes them.\\

\paragraph{Recursion.}
Inverting $u=G(\xi)\psi$ order by order fixes the kernels. Anchoring the inversion on
leg $2$ (so that $q_i\equiv|k_i|$, with $q_2$ in the denominator), they are most easily
built through the auxiliary kernels $E_n$ of Ref.~\cite{Ussembayev}, which satisfy,
with base case $E_3(1,2,3)=-\tfrac12\big(q_1q_2+k_1\!\cdot\!k_2\big)$,
\begin{equation}
E_N=\frac{q_2^{\,N-3}}{(N-2)!}\,E_3(1,2,3{+}\cdots{+}N)
-\sum_{m=1}^{N-3}\frac{q_2^{\,m}}{m!}\,E_{N-m}\big(1,2{+}\cdots{+}(m{+}2),m{+}3,\dots,N\big),
\label{eq:Erec}
\end{equation}
after which the Lagrangian kernels follow from
\begin{equation}
q_2\,F_N=\frac{2E_N}{q_1}
-\sum_{m=1}^{N-3}2E_{m+2}(-\Sigma_m,2,\dots,m{+}2)\,F_{N-m}(1,\Sigma_m,m{+}3,\dots,N),
\qquad \Sigma_m\equiv k_2{+}\cdots{+}k_{m+2}.
\label{eq:Frec}
\end{equation}
These are the routines \texttt{EKernelExpr} and \texttt{FKernelExpr} in the
repository~\cite{Hydrotopecode}. At cubic order the sum is empty, giving the base kernel
\begin{equation}
F_3=\frac{2E_3}{q_1q_2}=-1-\frac{k_1\!\cdot\!k_2}{q_1\,q_2}.
\label{eq:F3rec}
\end{equation}

\paragraph{Closed form and vertices.}
The recursion~\eqref{eq:Frec} resums into closed form. Writing the internal momenta
$P_\nu\equiv\sum_{i=2}^{\nu}k_i$ (so $P_2=k_2$ and $P_n=-k_1$ by momentum
conservation),
\begin{equation}
F_n=-\frac{1}{|k_2|}\sum_{\ell\ge0}\ \sum_{2=\nu_0<\nu_1<\cdots<\nu_\ell\le n}
\frac{|P_{\nu_\ell}|^{\,n-\nu_\ell}}{(n-\nu_\ell)!}
\prod_{j=1}^{\ell}\left[-\frac{1}{|P_{\nu_j}|}\,
\frac{|P_{\nu_{j-1}}|^{\,\nu_j-1-\nu_{j-1}}}{(\nu_j-1-\nu_{j-1})!}
\left(P_{\nu_{j-1}}\!\cdot k_{\nu_j}+\frac{|P_{\nu_{j-1}}|^2}{\nu_j-\nu_{j-1}}\right)\right],
\label{eq:Fclosed}
\end{equation}
where the vertices appearing in Eq.~\eqref{eq:Frec} satisfy $V_n=-\tfrac12 F_n$.
A kernel becomes the physical $n$-point vertex once
its two derivative legs are dressed with frequencies and all legs are symmetrized,
\begin{equation}
\mathcal{V}_n=-\frac{1}{2}\sum_{\sigma\in S_n}
\omega_{\sigma(1)}\,\omega_{\sigma(2)}\,F_n\big(k_{\sigma(1)},\dots,k_{\sigma(n)}\big) .
\label{eq:Vertex}
\end{equation}
This Lagrangian contact term $\mathcal{V}_n$ is what enters the Berends--Giele recursion.\\

\paragraph{Cubic and quartic.}
Evaluating \eqref{eq:Fclosed} at $n=3$ recovers \eqref{eq:F3rec}, whose dressing
$-\tfrac12\omega_1\omega_2F_3$ is the unsymmetrized cubic contact vertex of the main text \eqref{eq:surfLcubic}. At $n=4$
the four chains give, with $P_3=k_2+k_3$ and $P_4=-k_1$,
\begin{equation}
F_4=-\frac{1}{|k_2|}\left[\frac{|k_2|^2}{2}
-\big(k_2\!\cdot\!k_3+|k_2|^2\big)
-\frac{|k_2|}{|k_1|}\Big(k_2\!\cdot\!k_4+\frac{|k_2|^2}{2}\Big)
+\frac{\big(k_2\!\cdot\!k_3+|k_2|^2\big)\,P_3\!\cdot\!P_4}{|P_3|\,|P_4|}\right].
\label{eq:F4}
\end{equation}
for which the full Lagrangian contact term is $-\tfrac12\omega_1\omega_2F_4$.

\section{Recovering Lvov's five-wave matrix elements}
\label{sec:supp-lvov}

The first explicit one-dimensional five-wave amplitudes were obtained by
Lvov~\cite{Lvov}, who constructed the effective
five-wave Hamiltonian building on techniques developed in Ref.~\cite{DLZ95} by a canonical transformation that removed the four-wave interaction coefficients, and tabulated the symmetric matrix
element $T^{k_1k_2k_3}_{pq}$ on the resonant manifold
\begin{equation}
k_1+k_2+k_3=p+q,\qquad
\omega_{k_1}+\omega_{k_2}+\omega_{k_3}=\omega_p+\omega_q,\qquad
\omega_k=\sqrt{g|k|},
\label{eq:lvovmanifold}
\end{equation}
where every $\omega$ denotes a \emph{positive} dispersion value $\sqrt{g|k|}$
and the wave vectors $k_1,k_2,k_3,p,q$ carry independent signs. Lvov found that
for two sign orientations the matrix element vanishes identically, while the
remaining orientations give ``remarkably simple'' but conspicuously asymmetric and multi-branched expressions, and remarked that the
vanishing ``should have a deep physical meaning.'' We show here that the single hydrotope formula~\eqref{eq:hydrotope} reproduces his entire table:
the branches are the polygon regimes of $\mathcal{W}_5$, and the two vanishing
orientations are the one-minus sector.

\paragraph{Dictionary.}---We read Lvov's kinematics as five legs of $A_5$ with momenta $\{k_1,k_2,k_3,-p,-q\}$, summing to
zero by~\eqref{eq:lvovmanifold} and signed frequencies
$\{\omega_{k_1},\omega_{k_2},\omega_{k_3},-\omega_p,-\omega_q\}$, summing to
zero by energy conservation. Each leg's momentum sign is its $\sigma_i$, so the number of negative momenta fixes the sector, see Table~\ref{tab:sLvovsectors}. Comparing the two computations on the manifold
\eqref{eq:lvovmanifold} we find the exact relation (here $g=1$)
\begin{equation}
T^{k_1k_2k_3}_{pq}
=\frac{1}{16\,\pi^{3/2}}\,
\bigl(\omega_{k_1}\omega_{k_2}\omega_{k_3}\,\omega_p\,\omega_q\bigr)^{1/2}\,A_5
=\frac{1}{\pi^{3/2}}\,
\bigl(\omega_{k_1}\omega_{k_2}\omega_{k_3}\,\omega_p\,\omega_q\bigr)^{1/2}\,
\omega_a\omega_b\,\mathcal{P},
\label{eq:dictionary}
\end{equation}
where $\omega_a,\omega_b$ are the frequencies of the two minus legs,
$\mathcal{P}$ is the bracketed polytope value of Eq.~\eqref{eq:n5}
(so $A_5=16\,\omega_a\omega_b\,\mathcal{P}$), and the universal factor
$(\prod\omega)^{1/2}$ is the field-normalization Jacobian between Lvov's
canonical normal variable and the surface elevation $\xi$.

\begin{table}[h]
\renewcommand{\arraystretch}{1.15}
\begin{ruledtabular}
\begin{tabular}{clccl}
Config & Wave-vector signs $(k_1,k_2,k_3;p,q)$ & Neg.\  momenta & Our sector & $A_5$\\
\hline
(i)   & all positive                       & $2$ $(-p,-q)$ & two-minus                    & hydrotope\\
(ii)  & one $k<0$;\ \ $p,q>0$               & $3$           & three-minus $\equiv$ two-minus & hydrotope\\
(iii) & two $k<0$;\ \ $p,q>0$               & $4$           & one-plus                     & $0$\\
(iv)  & $k_{1,2,3}>0$;\ \ $p,q$ opposite    & $1$           & one-minus                    & $0$\\
(v)   & one $k<0$;\ \ $p,q$ opposite        & $2$           & two-minus                    & hydrotope\\
\end{tabular}
\end{ruledtabular}
\caption{Lvov's five sign orientations mapped to our convention in Eq.~\eqref{eq:conv}. At $n=5$ the three-minus configuration is the parity image
($k\to-k$) of the two-minus one, so (i), (ii), and (v) all give the hydrotope.
The two vanishing orientations (iii), (iv) are exactly the one-minus/one-plus
sector that vanishes.}
\label{tab:sLvovsectors}
\end{table}

For each nonzero orientation the two
minus legs carry $\beta=\min(|\omega_a|,|\omega_b|)$: legs $p,q$ in (i),
legs $k_1,k_2$ in (ii) (after the parity flip), and legs $k_3,p$ in (v). Lvov's
sub-branches are then distinguished by the ordering of the frequencies,
which is precisely the data selecting which corners of the box
$\prod[0,\omega_i^2]$ are clipped---i.e.\ the polygon regime of $\mathcal{W}_5$.
Table~\ref{tab:sLvovbranches} lists every branch of Lvov's Table~1 alongside the
regime it realizes; Eq.~\eqref{eq:dictionary} with the corresponding
$\mathcal{P}$ from Eq.~\eqref{eq:n5} reproduces each formula. Lvov's apparent
asymmetry is merely the labeling of the minimum leg and the ordering
$\omega_3^2\le \omega_4^2\le \omega_5^2$; the hydrotope is the single symmetric object beneath all seven branches.

\begin{table*}[h]
\renewcommand{\arraystretch}{1.4}
\begin{ruledtabular}
\begin{tabular}{cll c c}
Branch & Ordering & Lvov matrix element\ \ $\pi^{3/2}\,T^{k_1k_2k_3}_{pq}$\ \ ($g=1$) & Regime & Clips\\
\hline
(i)     & ---                                  & $2\,\omega_{k_1}^{5/2}\omega_{k_2}^{5/2}\omega_{k_3}^{5/2}\omega_p^{3/2}\omega_q^{3/2}\big/\max(\omega_{k_1}^2,\omega_{k_2}^2,\omega_{k_3}^2)$ & parallelogram & $2$\\
(ii\,a) & $\omega_{k_2}<\omega_{k_3}$          & $\omega_{k_1}^{3/2}\omega_{k_2}^{11/2}\omega_{k_3}^{1/2}\omega_p^{1/2}\omega_q^{1/2}$ & triangle & $0$\\
(ii\,b) & $\omega_{k_2}>\omega_{k_3}$          & $\omega_{k_1}^{3/2}\omega_{k_2}^{3/2}\omega_{k_3}^{5/2}\omega_p^{1/2}\omega_q^{1/2}\,(2\omega_{k_2}^2-\omega_{k_3}^2)$ & trapezoid & $1$\\
(v\,a)  & $\omega_{k_1}>\omega_{k_2}>\omega_{k_3}$ & $-\,\omega_{k_1}^{1/2}\omega_{k_2}^{1/2}\omega_{k_3}^{11/2}\omega_p^{3/2}\omega_q^{1/2}$ & triangle & $0$\\
(v\,b)  & $\omega_{k_1}>\omega_{k_3}>\omega_{k_2}$ & $\omega_{k_1}^{1/2}\omega_{k_2}^{5/2}\omega_{k_3}^{3/2}\omega_p^{3/2}\omega_q^{1/2}\,(\omega_{k_2}^2-2\omega_{k_3}^2)$ & trapezoid & $1$\\
(v\,c)  & $\omega_{k_3}>\omega_{k_1}>\omega_{k_2},\ \omega_p>\omega_q$ & $\omega_{k_1}^{1/2}\omega_{k_2}^{1/2}\omega_{k_3}^{3/2}\omega_p^{3/2}\omega_q^{1/2}\,(\omega_{k_1}^4+\omega_{k_2}^4-2\omega_{k_1}^2\omega_{k_3}^2-2\omega_{k_2}^2\omega_{k_3}^2+\omega_{k_3}^4)$ & pentagon & $2$\\
(v\,d)  & $\omega_{k_3}>\omega_{k_1}>\omega_{k_2},\ \omega_q>\omega_p$ & $-\,2\,\omega_{k_1}^{5/2}\omega_{k_2}^{5/2}\omega_{k_3}^{3/2}\omega_p^{3/2}\omega_q^{1/2}$ & parallelogram & $2$\\
\end{tabular}
\end{ruledtabular}
\caption{Every nonzero branch of Lvov's Table~1 and the hydrotope polygon
regime it realizes. The monomial branches are the triangle ($\mathcal{P}=\beta^4$)
and parallelogram ($\mathcal{P}=2a_3a_4$, where the largest plus-leg drops out, corresponding to Lvov's $1/\max$); the binomial brackets are the trapezoid
($\mathcal{P}=a_3(2\beta^2-a_3)$) and the degree-four bracket is the pentagon.
Relation~\eqref{eq:dictionary} with the matching $\mathcal{P}$ of
Eq.~\eqref{eq:n5} reproduces each entry.}
\label{tab:sLvovbranches}
\end{table*}

\newpage

\section{Visualisation of the six-point geometries}
\begin{figure*}[h]
\centering
\resizebox{0.82\textwidth}{!}{%
\begin{tikzpicture}
\draw[gray!45,line width=0.3pt,densely dotted] (-0.206,-0.275)--(-0.206,0.775);
\draw[gray!45,line width=0.3pt,densely dotted] (-0.206,-0.275)--(0.759,-0.457);
\draw[gray!45,line width=0.3pt,densely dotted] (-0.206,-0.275)--(-0.759,-0.593);
\draw[gray!55,line width=0.4pt] (-0.206,0.775)--(0.759,0.593);
\draw[gray!55,line width=0.4pt] (-0.206,0.775)--(-0.759,0.457);
\draw[gray!55,line width=0.4pt] (0.759,-0.457)--(0.759,0.593);
\draw[gray!55,line width=0.4pt] (0.759,-0.457)--(0.206,-0.775);
\draw[gray!55,line width=0.4pt] (0.759,0.593)--(0.206,0.275);
\draw[gray!55,line width=0.4pt] (-0.759,-0.593)--(-0.759,0.457);
\draw[gray!55,line width=0.4pt] (-0.759,-0.593)--(0.206,-0.775);
\draw[gray!55,line width=0.4pt] (-0.759,0.457)--(0.206,0.275);
\draw[gray!55,line width=0.4pt] (0.206,-0.775)--(0.206,0.275);
\draw[stedgeH] (-0.206,-0.275)--(-0.482,-0.434);
\draw[stedgeH] (-0.206,-0.275)--(0.276,-0.366);
\draw[stedgeH] (-0.206,-0.275)--(-0.206,0.250);
\fill[glass] (0.276,-0.366)--(-0.206,-0.275)--(-0.206,0.250)--cycle;
\fill[glass] (-0.206,0.250)--(-0.206,-0.275)--(-0.482,-0.434)--cycle;
\fill[glass] (-0.482,-0.434)--(-0.206,-0.275)--(0.276,-0.366)--cycle;
\fill[glass] (0.276,-0.366)--(-0.206,0.250)--(-0.482,-0.434)--cycle;
\draw[stedgeV] (0.276,-0.366)--(-0.482,-0.434);
\draw[stedgeV] (-0.206,0.250)--(-0.482,-0.434);
\draw[stedgeV] (-0.206,0.250)--(0.276,-0.366);
\draw[-{Latex},black,line width=0.35pt] (-0.206,-0.275)--(-0.759,-0.593);
\node[font=\scriptsize,below] at (-0.825,-0.631) {$t_3$};
\draw[-{Latex},black,line width=0.35pt] (-0.206,-0.275)--(0.759,-0.457);
\node[font=\scriptsize,right] at (0.875,-0.479) {$t_4$};
\draw[-{Latex},black,line width=0.35pt] (-0.206,-0.275)--(-0.206,0.775);
\node[font=\scriptsize,left] at (-0.206,0.901) {$t_5$};
\node[font=\small\bfseries] at (0.00,-1.24) {tetrahedron};
\node[font=\scriptsize,gray!45!black] at (0.00,-1.62) {$(4,6,4)$};
\draw[gray!45,line width=0.3pt,densely dotted] (3.148,-0.441)--(3.148,0.775);
\draw[gray!45,line width=0.3pt,densely dotted] (3.148,-0.441)--(4.266,-0.652);
\draw[gray!45,line width=0.3pt,densely dotted] (3.148,-0.441)--(2.934,-0.564);
\draw[gray!55,line width=0.4pt] (3.148,0.775)--(4.266,0.564);
\draw[gray!55,line width=0.4pt] (3.148,0.775)--(2.934,0.652);
\draw[gray!55,line width=0.4pt] (4.266,-0.652)--(4.266,0.564);
\draw[gray!55,line width=0.4pt] (4.266,-0.652)--(4.052,-0.775);
\draw[gray!55,line width=0.4pt] (4.266,0.564)--(4.052,0.441);
\draw[gray!55,line width=0.4pt] (2.934,-0.564)--(2.934,0.652);
\draw[gray!55,line width=0.4pt] (2.934,-0.564)--(4.052,-0.775);
\draw[gray!55,line width=0.4pt] (2.934,0.652)--(4.052,0.441);
\draw[gray!55,line width=0.4pt] (4.052,-0.775)--(4.052,0.441);
\draw[stedgeH] (3.148,-0.441)--(3.893,-0.582);
\draw[stedgeH] (3.148,-0.441)--(3.148,0.370);
\draw[stedgeH] (3.148,-0.441)--(2.934,-0.564);
\fill[glass] (3.893,-0.582)--(3.148,-0.441)--(3.148,0.370)--cycle;
\fill[glass] (3.148,0.370)--(3.148,-0.441)--(2.934,-0.564)--(2.934,-0.159)--cycle;
\fill[glass] (3.307,-0.634)--(2.934,-0.564)--(3.148,-0.441)--(3.893,-0.582)--cycle;
\fill[glass] (3.893,-0.582)--(3.148,0.370)--(2.934,-0.159)--(3.307,-0.634)--cycle;
\fill[glass] (2.934,-0.159)--(2.934,-0.564)--(3.307,-0.634)--cycle;
\draw[stedgeV] (3.148,0.370)--(3.893,-0.582);
\draw[stedgeV] (2.934,-0.564)--(2.934,-0.159);
\draw[stedgeV] (2.934,-0.564)--(3.307,-0.634);
\draw[stedgeV] (2.934,-0.159)--(3.307,-0.634);
\draw[stedgeV] (3.148,0.370)--(2.934,-0.159);
\draw[stedgeV] (3.893,-0.582)--(3.307,-0.634);
\node[font=\small\bfseries] at (3.60,-1.24) {triangular prism};
\node[font=\scriptsize,gray!45!black] at (3.60,-1.62) {$(6,9,5)$};
\draw[gray!45,line width=0.3pt,densely dotted] (7.077,-0.477)--(7.077,0.775);
\draw[gray!45,line width=0.3pt,densely dotted] (7.077,-0.477)--(7.652,-0.585);
\draw[gray!45,line width=0.3pt,densely dotted] (7.077,-0.477)--(6.748,-0.666);
\draw[gray!55,line width=0.4pt] (7.077,0.775)--(7.652,0.666);
\draw[gray!55,line width=0.4pt] (7.077,0.775)--(6.748,0.585);
\draw[gray!55,line width=0.4pt] (7.652,-0.585)--(7.652,0.666);
\draw[gray!55,line width=0.4pt] (7.652,-0.585)--(7.323,-0.775);
\draw[gray!55,line width=0.4pt] (7.652,0.666)--(7.323,0.477);
\draw[gray!55,line width=0.4pt] (6.748,-0.666)--(6.748,0.585);
\draw[gray!55,line width=0.4pt] (6.748,-0.666)--(7.323,-0.775);
\draw[gray!55,line width=0.4pt] (6.748,0.585)--(7.323,0.477);
\draw[gray!55,line width=0.4pt] (7.323,-0.775)--(7.323,0.477);
\draw[stedgeH] (7.077,-0.477)--(7.077,0.462);
\draw[stedgeH] (7.077,-0.477)--(6.748,-0.666);
\draw[stedgeH] (7.077,-0.477)--(7.652,-0.585);
\fill[glass] (7.652,-0.272)--(7.652,-0.585)--(7.077,-0.477)--(7.077,0.462)--cycle;
\fill[glass] (7.077,0.462)--(7.077,-0.477)--(6.748,-0.666)--(6.748,-0.353)--cycle;
\fill[glass] (7.035,-0.721)--(6.748,-0.666)--(7.077,-0.477)--(7.652,-0.585)--(7.488,-0.680)--cycle;
\fill[glass] (7.488,-0.680)--(7.652,-0.585)--(7.652,-0.272)--cycle;
\fill[glass] (7.488,-0.680)--(7.652,-0.272)--(7.077,0.462)--(6.748,-0.353)--(7.035,-0.721)--cycle;
\fill[glass] (6.748,-0.353)--(6.748,-0.666)--(7.035,-0.721)--cycle;
\draw[stedgeV] (6.748,-0.666)--(6.748,-0.353);
\draw[stedgeV] (6.748,-0.666)--(7.035,-0.721);
\draw[stedgeV] (6.748,-0.353)--(7.035,-0.721);
\draw[stedgeV] (7.652,-0.585)--(7.488,-0.680);
\draw[stedgeV] (7.652,-0.585)--(7.652,-0.272);
\draw[stedgeV] (7.652,-0.272)--(7.488,-0.680);
\draw[stedgeV] (7.077,0.462)--(6.748,-0.353);
\draw[stedgeV] (7.077,0.462)--(7.652,-0.272);
\draw[stedgeV] (7.035,-0.721)--(7.488,-0.680);
\node[font=\small\bfseries] at (7.20,-1.24) {8 vertices};
\node[font=\scriptsize,gray!45!black] at (7.20,-1.62) {$(8,12,6)$};
\draw[gray!45,line width=0.3pt,densely dotted] (10.732,-0.610)--(10.732,0.775);
\draw[gray!45,line width=0.3pt,densely dotted] (10.732,-0.610)--(11.050,-0.670);
\draw[gray!45,line width=0.3pt,densely dotted] (10.732,-0.610)--(10.550,-0.715);
\draw[gray!55,line width=0.4pt] (10.732,0.775)--(11.050,0.715);
\draw[gray!55,line width=0.4pt] (10.732,0.775)--(10.550,0.670);
\draw[gray!55,line width=0.4pt] (11.050,-0.670)--(11.050,0.715);
\draw[gray!55,line width=0.4pt] (11.050,-0.670)--(10.868,-0.775);
\draw[gray!55,line width=0.4pt] (11.050,0.715)--(10.868,0.610);
\draw[gray!55,line width=0.4pt] (10.550,-0.715)--(10.550,0.670);
\draw[gray!55,line width=0.4pt] (10.550,-0.715)--(10.868,-0.775);
\draw[gray!55,line width=0.4pt] (10.550,0.670)--(10.868,0.610);
\draw[gray!55,line width=0.4pt] (10.868,-0.775)--(10.868,0.610);
\draw[stedgeH] (10.732,-0.610)--(10.550,-0.715);
\draw[stedgeH] (10.732,-0.610)--(11.050,-0.670);
\draw[stedgeH] (10.732,-0.610)--(10.732,0.429);
\fill[glass] (10.868,-0.775)--(10.550,-0.715)--(10.732,-0.610)--(11.050,-0.670)--cycle;
\fill[glass] (11.050,0.022)--(11.050,-0.670)--(10.732,-0.610)--(10.732,0.429)--cycle;
\fill[glass] (10.732,0.429)--(10.732,-0.610)--(10.550,-0.715)--(10.550,-0.022)--cycle;
\fill[glass] (10.868,-0.429)--(10.868,-0.775)--(11.050,-0.670)--(11.050,0.022)--cycle;
\fill[glass] (10.550,-0.022)--(10.550,-0.715)--(10.868,-0.775)--(10.868,-0.429)--cycle;
\fill[glass] (10.868,-0.429)--(11.050,0.022)--(10.732,0.429)--(10.550,-0.022)--cycle;
\draw[stedgeV] (10.550,-0.715)--(10.868,-0.775);
\draw[stedgeV] (11.050,-0.670)--(10.868,-0.775);
\draw[stedgeV] (10.550,-0.715)--(10.550,-0.022);
\draw[stedgeV] (10.732,0.429)--(10.550,-0.022);
\draw[stedgeV] (10.868,-0.775)--(10.868,-0.429);
\draw[stedgeV] (10.550,-0.022)--(10.868,-0.429);
\draw[stedgeV] (11.050,-0.670)--(11.050,0.022);
\draw[stedgeV] (10.732,0.429)--(11.050,0.022);
\draw[stedgeV] (11.050,0.022)--(10.868,-0.429);
\node[font=\small\bfseries] at (10.80,-1.24) {parallelepiped};
\node[font=\scriptsize,gray!45!black] at (10.80,-1.62) {$(8,12,6)$};
\draw[gray!45,line width=0.3pt,densely dotted] (-0.206,-3.975)--(-0.206,-2.925);
\draw[gray!45,line width=0.3pt,densely dotted] (-0.206,-3.975)--(0.759,-4.157);
\draw[gray!45,line width=0.3pt,densely dotted] (-0.206,-3.975)--(-0.759,-4.293);
\draw[gray!55,line width=0.4pt] (-0.206,-2.925)--(0.759,-3.107);
\draw[gray!55,line width=0.4pt] (-0.206,-2.925)--(-0.759,-3.243);
\draw[gray!55,line width=0.4pt] (0.759,-4.157)--(0.759,-3.107);
\draw[gray!55,line width=0.4pt] (0.759,-4.157)--(0.206,-4.475);
\draw[gray!55,line width=0.4pt] (0.759,-3.107)--(0.206,-3.425);
\draw[gray!55,line width=0.4pt] (-0.759,-4.293)--(-0.759,-3.243);
\draw[gray!55,line width=0.4pt] (-0.759,-4.293)--(0.206,-4.475);
\draw[gray!55,line width=0.4pt] (-0.759,-3.243)--(0.206,-3.425);
\draw[gray!55,line width=0.4pt] (0.206,-4.475)--(0.206,-3.425);
\draw[stedgeH] (-0.206,-3.975)--(-0.759,-4.293);
\draw[stedgeH] (-0.206,-3.975)--(0.759,-4.157);
\draw[stedgeH] (-0.206,-3.975)--(-0.206,-2.925);
\fill[glass] (0.759,-3.632)--(0.759,-4.157)--(-0.206,-3.975)--(-0.206,-2.925)--(0.276,-3.016)--cycle;
\fill[glass] (0.276,-3.016)--(-0.206,-2.925)--(-0.482,-3.084)--cycle;
\fill[glass] (-0.482,-3.084)--(-0.206,-2.925)--(-0.206,-3.975)--(-0.759,-4.293)--(-0.759,-3.768)--cycle;
\fill[glass] (-0.276,-4.384)--(-0.759,-4.293)--(-0.206,-3.975)--(0.759,-4.157)--(0.482,-4.316)--cycle;
\fill[glass] (0.759,-3.632)--(0.482,-4.316)--(0.759,-4.157)--cycle;
\fill[glass] (-0.276,-4.384)--(0.482,-4.316)--(0.759,-3.632)--(0.276,-3.016)--(-0.482,-3.084)--(-0.759,-3.768)--cycle;
\fill[glass] (-0.759,-4.293)--(-0.276,-4.384)--(-0.759,-3.768)--cycle;
\draw[stedgeV] (-0.206,-2.925)--(0.276,-3.016);
\draw[stedgeV] (-0.206,-2.925)--(-0.482,-3.084);
\draw[stedgeV] (0.276,-3.016)--(-0.482,-3.084);
\draw[stedgeV] (0.759,-3.632)--(0.482,-4.316);
\draw[stedgeV] (0.759,-4.157)--(0.482,-4.316);
\draw[stedgeV] (0.759,-4.157)--(0.759,-3.632);
\draw[stedgeV] (-0.759,-4.293)--(-0.276,-4.384);
\draw[stedgeV] (-0.759,-3.768)--(-0.276,-4.384);
\draw[stedgeV] (-0.759,-4.293)--(-0.759,-3.768);
\draw[stedgeV] (-0.276,-4.384)--(0.482,-4.316);
\draw[stedgeV] (0.759,-3.632)--(0.276,-3.016);
\draw[stedgeV] (-0.759,-3.768)--(-0.482,-3.084);
\node[font=\small\bfseries] at (0.00,-4.94) {10 vertices};
\node[font=\scriptsize,gray!45!black] at (0.00,-5.32) {$(10,15,7)$};
\draw[gray!45,line width=0.3pt,densely dotted] (3.477,-4.177)--(3.477,-2.925);
\draw[gray!45,line width=0.3pt,densely dotted] (3.477,-4.177)--(4.052,-4.285);
\draw[gray!45,line width=0.3pt,densely dotted] (3.477,-4.177)--(3.148,-4.366);
\draw[gray!55,line width=0.4pt] (3.477,-2.925)--(4.052,-3.034);
\draw[gray!55,line width=0.4pt] (3.477,-2.925)--(3.148,-3.115);
\draw[gray!55,line width=0.4pt] (4.052,-4.285)--(4.052,-3.034);
\draw[gray!55,line width=0.4pt] (4.052,-4.285)--(3.723,-4.475);
\draw[gray!55,line width=0.4pt] (4.052,-3.034)--(3.723,-3.223);
\draw[gray!55,line width=0.4pt] (3.148,-4.366)--(3.148,-3.115);
\draw[gray!55,line width=0.4pt] (3.148,-4.366)--(3.723,-4.475);
\draw[gray!55,line width=0.4pt] (3.148,-3.115)--(3.723,-3.223);
\draw[gray!55,line width=0.4pt] (3.723,-4.475)--(3.723,-3.223);
\draw[stedgeH] (3.477,-4.177)--(3.148,-4.366);
\draw[stedgeH] (3.477,-4.177)--(4.052,-4.285);
\draw[stedgeH] (3.477,-4.177)--(3.477,-2.925);
\fill[glass] (4.052,-3.347)--(4.052,-4.285)--(3.477,-4.177)--(3.477,-2.925)--(3.765,-2.979)--cycle;
\fill[glass] (4.052,-4.285)--(3.723,-4.475)--(3.148,-4.366)--(3.477,-4.177)--cycle;
\fill[glass] (3.312,-3.020)--(3.477,-2.925)--(3.477,-4.177)--(3.148,-4.366)--(3.148,-3.428)--cycle;
\fill[glass] (3.765,-2.979)--(3.477,-2.925)--(3.312,-3.020)--cycle;
\fill[glass] (4.052,-3.347)--(3.723,-4.162)--(3.723,-4.475)--(4.052,-4.285)--cycle;
\fill[glass] (3.148,-3.428)--(3.723,-4.162)--(4.052,-3.347)--(3.765,-2.979)--(3.312,-3.020)--cycle;
\fill[glass] (3.148,-4.366)--(3.723,-4.475)--(3.723,-4.162)--(3.148,-3.428)--cycle;
\draw[stedgeV] (3.477,-2.925)--(3.765,-2.979);
\draw[stedgeV] (3.477,-2.925)--(3.312,-3.020);
\draw[stedgeV] (3.765,-2.979)--(3.312,-3.020);
\draw[stedgeV] (4.052,-4.285)--(3.723,-4.475);
\draw[stedgeV] (3.148,-4.366)--(3.723,-4.475);
\draw[stedgeV] (3.723,-4.475)--(3.723,-4.162);
\draw[stedgeV] (3.148,-3.428)--(3.723,-4.162);
\draw[stedgeV] (3.148,-4.366)--(3.148,-3.428);
\draw[stedgeV] (4.052,-3.347)--(3.723,-4.162);
\draw[stedgeV] (4.052,-4.285)--(4.052,-3.347);
\draw[stedgeV] (4.052,-3.347)--(3.765,-2.979);
\draw[stedgeV] (3.148,-3.428)--(3.312,-3.020);
\node[font=\small\bfseries] at (3.60,-4.94) {10 vertices};
\node[font=\scriptsize,gray!45!black] at (3.60,-5.32) {$(10,15,7)$};
\draw[gray!45,line width=0.3pt,densely dotted] (6.794,-4.110)--(6.794,-2.925);
\draw[gray!45,line width=0.3pt,densely dotted] (6.794,-4.110)--(7.883,-4.315);
\draw[gray!45,line width=0.3pt,densely dotted] (6.794,-4.110)--(6.517,-4.269);
\draw[gray!55,line width=0.4pt] (6.794,-2.925)--(7.883,-3.131);
\draw[gray!55,line width=0.4pt] (6.794,-2.925)--(6.517,-3.085);
\draw[gray!55,line width=0.4pt] (7.883,-4.315)--(7.883,-3.131);
\draw[gray!55,line width=0.4pt] (7.883,-4.315)--(7.606,-4.475);
\draw[gray!55,line width=0.4pt] (7.883,-3.131)--(7.606,-3.290);
\draw[gray!55,line width=0.4pt] (6.517,-4.269)--(6.517,-3.085);
\draw[gray!55,line width=0.4pt] (6.517,-4.269)--(7.606,-4.475);
\draw[gray!55,line width=0.4pt] (6.517,-3.085)--(7.606,-3.290);
\draw[gray!55,line width=0.4pt] (7.606,-4.475)--(7.606,-3.290);
\draw[stedgeH] (6.794,-4.110)--(6.517,-4.269);
\draw[stedgeH] (6.794,-4.110)--(7.883,-4.315);
\draw[stedgeH] (6.794,-4.110)--(6.794,-2.925);
\fill[glass] (6.794,-2.925)--(6.794,-4.110)--(6.517,-4.269)--(6.517,-3.085)--cycle;
\fill[glass] (7.606,-4.475)--(6.517,-4.269)--(6.794,-4.110)--(7.883,-4.315)--cycle;
\fill[glass] (7.883,-3.460)--(7.883,-4.315)--(6.794,-4.110)--(6.794,-2.925)--(7.581,-3.074)--cycle;
\fill[glass] (6.794,-2.925)--(6.517,-3.085)--(6.819,-3.142)--(7.581,-3.074)--cycle;
\fill[glass] (6.819,-3.142)--(6.517,-3.085)--(6.517,-4.269)--(7.606,-4.475)--(7.606,-4.146)--cycle;
\fill[glass] (7.883,-3.460)--(7.606,-4.146)--(7.606,-4.475)--(7.883,-4.315)--cycle;
\fill[glass] (6.819,-3.142)--(7.606,-4.146)--(7.883,-3.460)--(7.581,-3.074)--cycle;
\draw[stedgeV] (6.517,-4.269)--(7.606,-4.475);
\draw[stedgeV] (7.883,-4.315)--(7.606,-4.475);
\draw[stedgeV] (6.517,-4.269)--(6.517,-3.085);
\draw[stedgeV] (6.794,-2.925)--(6.517,-3.085);
\draw[stedgeV] (7.883,-4.315)--(7.883,-3.460);
\draw[stedgeV] (6.794,-2.925)--(7.581,-3.074);
\draw[stedgeV] (7.883,-3.460)--(7.581,-3.074);
\draw[stedgeV] (7.883,-3.460)--(7.606,-4.146);
\draw[stedgeV] (7.606,-4.475)--(7.606,-4.146);
\draw[stedgeV] (6.517,-3.085)--(6.819,-3.142);
\draw[stedgeV] (7.606,-4.146)--(6.819,-3.142);
\draw[stedgeV] (7.581,-3.074)--(6.819,-3.142);
\node[font=\small\bfseries] at (7.20,-4.94) {10 vertices};
\node[font=\scriptsize,gray!45!black] at (7.20,-5.32) {$(10,15,7)$};
\draw[gray!45,line width=0.3pt,densely dotted] (10.594,-3.975)--(10.594,-2.925);
\draw[gray!45,line width=0.3pt,densely dotted] (10.594,-3.975)--(11.559,-4.157);
\draw[gray!45,line width=0.3pt,densely dotted] (10.594,-3.975)--(10.041,-4.293);
\draw[gray!55,line width=0.4pt] (10.594,-2.925)--(11.559,-3.107);
\draw[gray!55,line width=0.4pt] (10.594,-2.925)--(10.041,-3.243);
\draw[gray!55,line width=0.4pt] (11.559,-4.157)--(11.559,-3.107);
\draw[gray!55,line width=0.4pt] (11.559,-4.157)--(11.006,-4.475);
\draw[gray!55,line width=0.4pt] (11.559,-3.107)--(11.006,-3.425);
\draw[gray!55,line width=0.4pt] (10.041,-4.293)--(10.041,-3.243);
\draw[gray!55,line width=0.4pt] (10.041,-4.293)--(11.006,-4.475);
\draw[gray!55,line width=0.4pt] (10.041,-3.243)--(11.006,-3.425);
\draw[gray!55,line width=0.4pt] (11.006,-4.475)--(11.006,-3.425);
\draw[stedgeH] (10.594,-3.450)--(10.318,-4.134);
\draw[stedgeH] (10.594,-3.450)--(11.076,-4.066);
\draw[stedgeH] (11.076,-4.066)--(10.318,-4.134);
\draw[stedgeH] (10.594,-2.925)--(10.594,-3.450);
\draw[stedgeH] (11.559,-4.157)--(11.076,-4.066);
\draw[stedgeH] (10.041,-4.293)--(10.318,-4.134);
\fill[glass] (10.318,-4.134)--(10.594,-3.450)--(11.076,-4.066)--cycle;
\fill[glass] (11.076,-4.066)--(10.594,-3.450)--(10.594,-2.925)--(11.076,-3.016)--(11.559,-3.632)--(11.559,-4.157)--cycle;
\fill[glass] (11.076,-3.016)--(10.594,-2.925)--(10.318,-3.084)--cycle;
\fill[glass] (10.041,-3.768)--(10.318,-3.084)--(10.594,-2.925)--(10.594,-3.450)--(10.318,-4.134)--(10.041,-4.293)--cycle;
\fill[glass] (10.318,-4.134)--(11.076,-4.066)--(11.559,-4.157)--(11.282,-4.316)--(10.524,-4.384)--(10.041,-4.293)--cycle;
\fill[glass] (11.282,-4.316)--(11.559,-4.157)--(11.559,-3.632)--cycle;
\fill[glass] (10.524,-4.384)--(11.282,-4.316)--(11.559,-3.632)--(11.076,-3.016)--(10.318,-3.084)--(10.041,-3.768)--cycle;
\fill[glass] (10.524,-4.384)--(10.041,-3.768)--(10.041,-4.293)--cycle;
\draw[stedgeV] (10.041,-3.768)--(10.524,-4.384);
\draw[stedgeV] (10.041,-4.293)--(10.041,-3.768);
\draw[stedgeV] (10.041,-4.293)--(10.524,-4.384);
\draw[stedgeV] (10.594,-2.925)--(11.076,-3.016);
\draw[stedgeV] (10.594,-2.925)--(10.318,-3.084);
\draw[stedgeV] (11.076,-3.016)--(10.318,-3.084);
\draw[stedgeV] (11.559,-4.157)--(11.282,-4.316);
\draw[stedgeV] (11.559,-4.157)--(11.559,-3.632);
\draw[stedgeV] (11.559,-3.632)--(11.282,-4.316);
\draw[stedgeV] (11.559,-3.632)--(11.076,-3.016);
\draw[stedgeV] (10.041,-3.768)--(10.318,-3.084);
\draw[stedgeV] (10.524,-4.384)--(11.282,-4.316);
\node[font=\small\bfseries] at (10.80,-4.94) {trunc.\ tetrahedron};
\node[font=\scriptsize,gray!45!black] at (10.80,-5.32) {$(12,18,8)$};
\draw[gray!45,line width=0.3pt,densely dotted] (-0.123,-7.877)--(-0.123,-6.625);
\draw[gray!45,line width=0.3pt,densely dotted] (-0.123,-7.877)--(0.452,-7.985);
\draw[gray!45,line width=0.3pt,densely dotted] (-0.123,-7.877)--(-0.452,-8.066);
\draw[gray!55,line width=0.4pt] (-0.123,-6.625)--(0.452,-6.734);
\draw[gray!55,line width=0.4pt] (-0.123,-6.625)--(-0.452,-6.815);
\draw[gray!55,line width=0.4pt] (0.452,-7.985)--(0.452,-6.734);
\draw[gray!55,line width=0.4pt] (0.452,-7.985)--(0.123,-8.175);
\draw[gray!55,line width=0.4pt] (0.452,-6.734)--(0.123,-6.923);
\draw[gray!55,line width=0.4pt] (-0.452,-8.066)--(-0.452,-6.815);
\draw[gray!55,line width=0.4pt] (-0.452,-8.066)--(0.123,-8.175);
\draw[gray!55,line width=0.4pt] (-0.452,-6.815)--(0.123,-6.923);
\draw[gray!55,line width=0.4pt] (0.123,-8.175)--(0.123,-6.923);
\draw[stedgeH] (-0.123,-7.564)--(-0.288,-7.972);
\draw[stedgeH] (-0.123,-7.564)--(0.165,-7.931);
\draw[stedgeH] (0.165,-7.931)--(-0.288,-7.972);
\draw[stedgeH] (0.452,-7.985)--(0.165,-7.931);
\draw[stedgeH] (-0.452,-8.066)--(-0.288,-7.972);
\draw[stedgeH] (-0.123,-6.625)--(-0.123,-7.564);
\fill[glass] (-0.288,-7.972)--(-0.123,-7.564)--(0.165,-7.931)--cycle;
\fill[glass] (0.165,-7.931)--(-0.123,-7.564)--(-0.123,-6.625)--(0.165,-6.679)--(0.452,-7.047)--(0.452,-7.985)--cycle;
\fill[glass] (-0.288,-7.972)--(0.165,-7.931)--(0.452,-7.985)--(0.123,-8.175)--(-0.452,-8.066)--cycle;
\fill[glass] (-0.452,-8.066)--(-0.452,-7.128)--(-0.288,-6.720)--(-0.123,-6.625)--(-0.123,-7.564)--(-0.288,-7.972)--cycle;
\fill[glass] (0.165,-6.679)--(-0.123,-6.625)--(-0.288,-6.720)--cycle;
\fill[glass] (0.452,-7.047)--(0.123,-7.862)--(0.123,-8.175)--(0.452,-7.985)--cycle;
\fill[glass] (0.123,-7.862)--(0.452,-7.047)--(0.165,-6.679)--(-0.288,-6.720)--(-0.452,-7.128)--cycle;
\fill[glass] (0.123,-8.175)--(0.123,-7.862)--(-0.452,-7.128)--(-0.452,-8.066)--cycle;
\draw[stedgeV] (-0.123,-6.625)--(0.165,-6.679);
\draw[stedgeV] (-0.123,-6.625)--(-0.288,-6.720);
\draw[stedgeV] (0.165,-6.679)--(-0.288,-6.720);
\draw[stedgeV] (0.123,-8.175)--(0.123,-7.862);
\draw[stedgeV] (-0.452,-7.128)--(0.123,-7.862);
\draw[stedgeV] (-0.452,-8.066)--(-0.452,-7.128);
\draw[stedgeV] (-0.452,-8.066)--(0.123,-8.175);
\draw[stedgeV] (0.452,-7.047)--(0.123,-7.862);
\draw[stedgeV] (0.452,-7.985)--(0.123,-8.175);
\draw[stedgeV] (0.452,-7.985)--(0.452,-7.047);
\draw[stedgeV] (0.452,-7.047)--(0.165,-6.679);
\draw[stedgeV] (-0.452,-7.128)--(-0.288,-6.720);
\node[font=\small\bfseries] at (0.00,-8.64) {12 vertices};
\node[font=\scriptsize,gray!45!black] at (0.00,-9.02) {$(12,18,8)$};
\draw[gray!45,line width=0.3pt,densely dotted] (3.394,-7.675)--(3.394,-6.625);
\draw[gray!45,line width=0.3pt,densely dotted] (3.394,-7.675)--(4.359,-7.857);
\draw[gray!45,line width=0.3pt,densely dotted] (3.394,-7.675)--(2.841,-7.993);
\draw[gray!55,line width=0.4pt] (3.394,-6.625)--(4.359,-6.807);
\draw[gray!55,line width=0.4pt] (3.394,-6.625)--(2.841,-6.943);
\draw[gray!55,line width=0.4pt] (4.359,-7.857)--(4.359,-6.807);
\draw[gray!55,line width=0.4pt] (4.359,-7.857)--(3.806,-8.175);
\draw[gray!55,line width=0.4pt] (4.359,-6.807)--(3.806,-7.125);
\draw[gray!55,line width=0.4pt] (2.841,-7.993)--(2.841,-6.943);
\draw[gray!55,line width=0.4pt] (2.841,-7.993)--(3.806,-8.175);
\draw[gray!55,line width=0.4pt] (2.841,-6.943)--(3.806,-7.125);
\draw[gray!55,line width=0.4pt] (3.806,-8.175)--(3.806,-7.125);
\draw[stedgeH] (3.394,-7.150)--(3.118,-7.834);
\draw[stedgeH] (3.394,-7.150)--(3.876,-7.766);
\draw[stedgeH] (3.876,-7.766)--(3.118,-7.834);
\draw[stedgeH] (3.394,-6.625)--(3.394,-7.150);
\draw[stedgeH] (4.359,-7.857)--(3.876,-7.766);
\draw[stedgeH] (2.841,-7.993)--(3.118,-7.834);
\fill[glass] (3.118,-7.834)--(3.394,-7.150)--(3.876,-7.766)--cycle;
\fill[glass] (3.876,-7.766)--(3.394,-7.150)--(3.394,-6.625)--(4.359,-6.807)--(4.359,-7.857)--cycle;
\fill[glass] (2.841,-7.993)--(2.841,-6.943)--(3.394,-6.625)--(3.394,-7.150)--(3.118,-7.834)--cycle;
\fill[glass] (2.841,-7.993)--(3.118,-7.834)--(3.876,-7.766)--(4.359,-7.857)--(3.806,-8.175)--cycle;
\fill[glass] (4.082,-6.966)--(4.359,-6.807)--(3.394,-6.625)--(2.841,-6.943)--(3.324,-7.034)--cycle;
\fill[glass] (3.806,-7.650)--(3.806,-8.175)--(4.359,-7.857)--(4.359,-6.807)--(4.082,-6.966)--cycle;
\fill[glass] (3.324,-7.034)--(2.841,-6.943)--(2.841,-7.993)--(3.806,-8.175)--(3.806,-7.650)--cycle;
\fill[glass] (3.324,-7.034)--(3.806,-7.650)--(4.082,-6.966)--cycle;
\draw[stedgeV] (3.806,-7.650)--(3.324,-7.034);
\draw[stedgeV] (3.806,-7.650)--(4.082,-6.966);
\draw[stedgeV] (3.324,-7.034)--(4.082,-6.966);
\draw[stedgeV] (3.394,-6.625)--(4.359,-6.807);
\draw[stedgeV] (4.359,-7.857)--(4.359,-6.807);
\draw[stedgeV] (2.841,-7.993)--(2.841,-6.943);
\draw[stedgeV] (3.394,-6.625)--(2.841,-6.943);
\draw[stedgeV] (4.359,-7.857)--(3.806,-8.175);
\draw[stedgeV] (2.841,-7.993)--(3.806,-8.175);
\draw[stedgeV] (3.806,-8.175)--(3.806,-7.650);
\draw[stedgeV] (4.359,-6.807)--(4.082,-6.966);
\draw[stedgeV] (2.841,-6.943)--(3.324,-7.034);
\node[font=\small\bfseries] at (3.60,-8.64) {12 vertices};
\node[font=\scriptsize,gray!45!black] at (3.60,-9.02) {$(12,18,8)$};
\draw[gray!45,line width=0.3pt,densely dotted] (6.817,-7.795)--(6.817,-6.625);
\draw[gray!45,line width=0.3pt,densely dotted] (6.817,-7.795)--(7.891,-7.998);
\draw[gray!45,line width=0.3pt,densely dotted] (6.817,-7.795)--(6.509,-7.972);
\draw[gray!55,line width=0.4pt] (6.817,-6.625)--(7.891,-6.828);
\draw[gray!55,line width=0.4pt] (6.817,-6.625)--(6.509,-6.802);
\draw[gray!55,line width=0.4pt] (7.891,-7.998)--(7.891,-6.828);
\draw[gray!55,line width=0.4pt] (7.891,-7.998)--(7.583,-8.175);
\draw[gray!55,line width=0.4pt] (7.891,-6.828)--(7.583,-7.005);
\draw[gray!55,line width=0.4pt] (6.509,-7.972)--(6.509,-6.802);
\draw[gray!55,line width=0.4pt] (6.509,-7.972)--(7.583,-8.175);
\draw[gray!55,line width=0.4pt] (6.509,-6.802)--(7.583,-7.005);
\draw[gray!55,line width=0.4pt] (7.583,-8.175)--(7.583,-7.005);
\draw[stedgeH] (6.817,-7.502)--(6.663,-7.883);
\draw[stedgeH] (6.817,-7.502)--(7.085,-7.846);
\draw[stedgeH] (7.085,-7.846)--(6.663,-7.883);
\draw[stedgeH] (6.817,-6.625)--(6.817,-7.502);
\draw[stedgeH] (6.509,-7.972)--(6.663,-7.883);
\draw[stedgeH] (7.891,-7.998)--(7.085,-7.846);
\fill[glass] (6.663,-7.883)--(6.817,-7.502)--(7.085,-7.846)--cycle;
\fill[glass] (6.509,-7.972)--(6.509,-6.802)--(6.817,-6.625)--(6.817,-7.502)--(6.663,-7.883)--cycle;
\fill[glass] (7.085,-7.846)--(6.817,-7.502)--(6.817,-6.625)--(7.623,-6.777)--(7.891,-7.121)--(7.891,-7.998)--cycle;
\fill[glass] (6.663,-7.883)--(7.085,-7.846)--(7.891,-7.998)--(7.583,-8.175)--(6.509,-7.972)--cycle;
\fill[glass] (6.817,-6.625)--(6.509,-6.802)--(6.777,-6.853)--(7.623,-6.777)--cycle;
\fill[glass] (7.891,-7.121)--(7.583,-7.883)--(7.583,-8.175)--(7.891,-7.998)--cycle;
\fill[glass] (6.777,-6.853)--(6.509,-6.802)--(6.509,-7.972)--(7.583,-8.175)--(7.583,-7.883)--cycle;
\fill[glass] (6.777,-6.853)--(7.583,-7.883)--(7.891,-7.121)--(7.623,-6.777)--cycle;
\draw[stedgeV] (6.509,-7.972)--(6.509,-6.802);
\draw[stedgeV] (6.817,-6.625)--(6.509,-6.802);
\draw[stedgeV] (7.891,-7.998)--(7.583,-8.175);
\draw[stedgeV] (6.509,-7.972)--(7.583,-8.175);
\draw[stedgeV] (6.817,-6.625)--(7.623,-6.777);
\draw[stedgeV] (7.891,-7.121)--(7.623,-6.777);
\draw[stedgeV] (7.891,-7.998)--(7.891,-7.121);
\draw[stedgeV] (7.891,-7.121)--(7.583,-7.883);
\draw[stedgeV] (7.583,-8.175)--(7.583,-7.883);
\draw[stedgeV] (6.509,-6.802)--(6.777,-6.853);
\draw[stedgeV] (7.583,-7.883)--(6.777,-6.853);
\draw[stedgeV] (7.623,-6.777)--(6.777,-6.853);
\node[font=\small\bfseries] at (7.20,-8.64) {12 vertices};
\node[font=\scriptsize,gray!45!black] at (7.20,-9.02) {$(12,18,8)$};
\draw[gray!45,line width=0.3pt,densely dotted] (10.348,-7.841)--(10.348,-6.625);
\draw[gray!45,line width=0.3pt,densely dotted] (10.348,-7.841)--(11.466,-8.052);
\draw[gray!45,line width=0.3pt,densely dotted] (10.348,-7.841)--(10.134,-7.964);
\draw[gray!55,line width=0.4pt] (10.348,-6.625)--(11.466,-6.836);
\draw[gray!55,line width=0.4pt] (10.348,-6.625)--(10.134,-6.748);
\draw[gray!55,line width=0.4pt] (11.466,-8.052)--(11.466,-6.836);
\draw[gray!55,line width=0.4pt] (11.466,-8.052)--(11.252,-8.175);
\draw[gray!55,line width=0.4pt] (11.466,-6.836)--(11.252,-6.959);
\draw[gray!55,line width=0.4pt] (10.134,-7.964)--(10.134,-6.748);
\draw[gray!55,line width=0.4pt] (10.134,-7.964)--(11.252,-8.175);
\draw[gray!55,line width=0.4pt] (10.134,-6.748)--(11.252,-6.959);
\draw[gray!55,line width=0.4pt] (11.252,-8.175)--(11.252,-6.959);
\draw[stedgeH] (10.348,-6.625)--(10.348,-7.030);
\draw[stedgeH] (10.348,-7.030)--(10.134,-7.559);
\draw[stedgeH] (10.348,-7.030)--(11.093,-7.982);
\draw[stedgeH] (11.093,-7.982)--(10.507,-8.034);
\draw[stedgeH] (11.466,-8.052)--(11.093,-7.982);
\fill[glass] (10.507,-8.034)--(10.134,-7.559)--(10.348,-7.030)--(11.093,-7.982)--cycle;
\fill[glass] (10.134,-6.748)--(10.348,-6.625)--(10.348,-7.030)--(10.134,-7.559)--cycle;
\fill[glass] (11.466,-8.052)--(11.093,-7.982)--(10.348,-7.030)--(10.348,-6.625)--(11.093,-6.766)--(11.466,-7.241)--cycle;
\fill[glass] (11.093,-7.982)--(11.466,-8.052)--(11.252,-8.175)--(10.507,-8.034)--cycle;
\fill[glass] (11.093,-6.766)--(10.348,-6.625)--(10.134,-6.748)--(10.507,-6.818)--cycle;
\fill[glass] (11.252,-8.175)--(11.252,-7.770)--(10.507,-6.818)--(10.134,-6.748)--(10.134,-7.559)--(10.507,-8.034)--cycle;
\fill[glass] (11.466,-7.241)--(11.252,-7.770)--(11.252,-8.175)--(11.466,-8.052)--cycle;
\fill[glass] (10.507,-6.818)--(11.252,-7.770)--(11.466,-7.241)--(11.093,-6.766)--cycle;
\draw[stedgeV] (10.348,-6.625)--(10.134,-6.748);
\draw[stedgeV] (10.134,-6.748)--(10.134,-7.559);
\draw[stedgeV] (10.134,-7.559)--(10.507,-8.034);
\draw[stedgeV] (10.348,-6.625)--(11.093,-6.766);
\draw[stedgeV] (11.466,-7.241)--(11.093,-6.766);
\draw[stedgeV] (11.466,-8.052)--(11.466,-7.241);
\draw[stedgeV] (10.134,-6.748)--(10.507,-6.818);
\draw[stedgeV] (11.093,-6.766)--(10.507,-6.818);
\draw[stedgeV] (11.252,-7.770)--(10.507,-6.818);
\draw[stedgeV] (11.466,-7.241)--(11.252,-7.770);
\draw[stedgeV] (11.252,-8.175)--(11.252,-7.770);
\draw[stedgeV] (11.252,-8.175)--(10.507,-8.034);
\draw[stedgeV] (11.466,-8.052)--(11.252,-8.175);
\node[font=\small\bfseries] at (10.80,-8.64) {12 vertices};
\node[font=\scriptsize,gray!45!black] at (10.80,-9.02) {$(12,18,8)$};\end{tikzpicture}}
\caption{All twelve combinatorially distinct six-wave geometries. Each is the hydrotope
$\mathcal{W}_6$ realized as the box $[0,\omega_3^2]\times[0,\omega_4^2]\times[0,\omega_5^2]$
in $(t_3,t_4,t_5)$ cut by the slab $\beta^2-\omega_6^2\le t_3+t_4+t_5\le\beta^2$
(eliminating $t_6$; ordering $\omega_3^2\le\cdots\le\omega_6^2$). The geometric objects are labeled by the number of vertices, edges and faces $(V,E,F)$.}
\label{fig:s6}
\end{figure*}
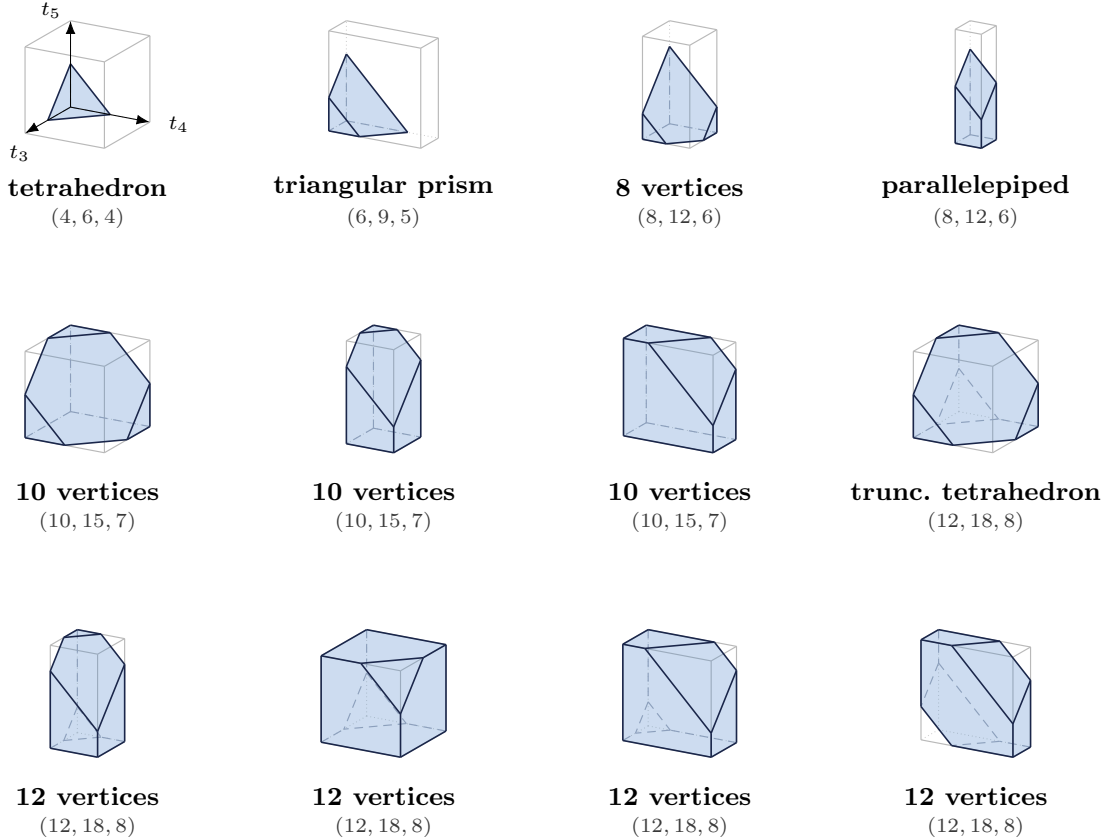

Going beyond five point, we show in Fig.~\ref{fig:s6} the geometric sectors that are realized in the six point amplitude.

\end{document}